\documentclass[floats,floatfix,showpacs,amssymb,prd,twocolumn,superscriptaddress,nofootinbib]{revtex4-1}

\usepackage{amssymb,amsmath,verbatim,mathtools,needspace,enumitem,etoolbox,graphicx,physics,microtype}

\usepackage[dvipsnames, usenames]{xcolor}

\definecolor{linkcolor}{rgb}{0.0,0.3,0.5}
\usepackage[unicode, colorlinks=true, linkcolor=linkcolor, citecolor=linkcolor, filecolor=linkcolor,urlcolor=linkcolor, pdfusetitle]{hyperref}
\usepackage[all]{hypcap}
\usepackage[T1]{fontenc}
\usepackage[utf8]{inputenc}

\allowdisplaybreaks
\def\araa{\rm{ARA\&A}}
\def\apj{\rm{ApJ}}
\def\apjl{\rm{ApJ}}
\def\apjs{\rm{ApJS}}
\def\aap{\rm{A\&A}}

\def\mnras{\rm{MNRAS}}
\def\prd{\rm{PRD}}
\def\prl{\rm{PRL}}
\def\prx{\rm{PRX}}

\def\nat{\rm{Nature}}

\def\cqg{\rm{CQG}}
\def\lrr{\rm{LRR}}
\def\bain{\rm{Bull.~Astron.~Inst.~Netherlands}}

\newcommand{\caltech}{\affiliation{TAPIR 350-17, California Institute of Technology, 1200 E California Boulevard, Pasadena, CA 91125, USA}}
\newcommand{\rochester}{\affiliation{Center for Computational Relativity and Gravitation, Rochester Institute of Technology,
85 Lomb Memorial Drive, Rochester, NY 14623, USA}}
\newcommand{\olemiss}{\affiliation{Department of Physics and Astronomy, The University of Mississippi, University, MS 38677, USA}}

\newcommand{\jhu}{\affiliation{Department of Physics and Astronomy, Johns Hopkins University, 3400 N. Charles
Street, Baltimore, MD 21218, USA}}
\newcommand{\dallas}{\affiliation{Department of Physics, The University of Texas at Dallas, Richardson, Texas 75080, USA}}
\newcommand{\warsaw}{\affiliation{Nicolaus Copernicus Astronomical Center, Polish Academy of Sciences, ul. Bartycka 18, 00-716 Warsaw, Poland}}

\begin{document}

\title{Spin orientations of merging black holes formed from the evolution of stellar binaries}

\author{Davide Gerosa}
\thanks{Einstein Fellow}
\email{dgerosa@caltech.edu}
\caltech
\author{Emanuele Berti} \jhu \olemiss
\author{Richard O'Shaughnessy} \rochester
\author{\\Krzysztof Belczynski} \warsaw
\author{Michael Kesden} \dallas
\author{Daniel Wysocki} \rochester
\author{Wojciech Gladysz} \warsaw
\pacs{}

\date{\today}

\begin{abstract}

We study the expected spin misalignments of merging binary black holes formed in isolation by combining state-of-the-art population-synthesis models with efficient post-Newtonian evolutions, thus tracking sources from stellar formation to gravitational-wave detection. We present extensive predictions of the properties of sources detectable by both current and future interferometers. We account for the fact that detectors are more sensitive to spinning black-hole binaries with suitable spin orientations and find that this significantly impacts the population of sources detectable by LIGO, while this is not the case for third-generation detectors. We find that three formation pathways, differentiated by the order of core collapse and common-envelope phases, dominate the observed population, and that their relative importance critically depends on the recoils imparted to black holes at birth. Our models suggest that measurements of the 
``effective-spin'' parameter $\chi_{\rm eff}$ will allow for powerful constraints. For instance, we find that the role of spin magnitudes and spin directions in $\chi_{\rm eff}$ can be %
largely
 disentangled, and that the symmetry of the effective-spin distribution is a robust indicator of the binary's formation history. Our predictions for individual spin directions and their precessional morphologies confirm and extend early toy models, while exploring substantially more realistic and broader sets of initial conditions.  Our main conclusion is that specific subpopulations of black-hole binaries will exhibit distinctive precessional dynamics: these classes include (but are not limited to) sources where stellar tidal interactions act on sufficiently short timescales, and massive binaries produced in pulsational pair-instability supernovae. Measurements of black-hole spin orientations have enormous potential to constrain specific evolutionary processes in the lives of massive binary stars.

\end{abstract}

\maketitle

\section{Introduction}

Gravitational-wave (GW) observations of merging black-hole (BH) binaries have the potential to unveil the fate of massive stars. As they exhaust all the available fuel, stars with initial masses $M\gtrsim 8 M_\odot$ are  expected to undergo gravitational collapse. About $\sim 15\%$ of them are predicted to form BHs \cite{2011ApJ...730...70O}.
The detection of stellar-origin BHs in a binary system requires not only the formation of BHs in the first place, but also the occurrence of astrophysical processes that can dissipate enough energy and angular momentum to bring the orbital separation below  $r\sim 50 R_\odot$, where GW damping can drive the binary to merger \cite{1964PhRv..136.1224P}.

There are two main classes of formation models, depending on whether  (i) the two BHs spend their entire lives together as stars, or (ii) they form separately and meet later. In models belonging to class (i), BH binaries are the end product of the life of  binaries of massive stars \cite{2014LRR....17....3P}. Each of the two stars undergoes gravitational collapse and, if the binary is not disrupted, a binary BH is left behind. A common-envelope phase -- where the core/remnant of one the two objects sinks into the outer layers of its companion \cite{1976IAUS...73...75P} -- is typically invoked to dissipate enough angular momentum and produce a merging binary. %
Models of class (ii) instead require dense stellar environments to facilitate the assembly of multiple BHs and many-body interactions to harden the binary \cite{2013LRR....16....4B}.  For a comprehensive review on BH-binary formation channels see, e.g., \cite{2018arXiv180605820M,2016ApJ...818L..22A} and references therein.

The most obvious observable to confirm or rule out formation channels is the merger rate, currently constrained to the range $12 - 213\;{\rm Gpc}^{-3} {\rm yr}^{-1}$ \cite{2017PhRvL.118v1101A}. These large uncertainties leave ample room for models in both classes to match the observational constraints which, at present, do not allow us to confirm or rule out any of the preferred scenarios.

Measurements of the BH masses also tend to be poorly constraining, partly because of a selection bias: more massive systems are visible farther out. This tends to wash out differences in the intrinsic distributions, such that the observable distributions predicted by various models all tend to overlap. In practice, $\mathcal{O}(100)$ observations could be necessary before strong constraints can be placed using mass measurements  \cite{2015ApJ...810...58S,2017ApJ...846...82Z,2017PhRvD..95j3010K,2018MNRAS.477.4685B} (although sharp features like a mass cutoff will be accessible earlier \cite{2017ApJ...851L..25F,2018ApJ...856..173T,2018arXiv180506442W}).

Merger redshifts are also weak observables. They are expected to be set by the star formation history \cite{2014ARA&A..52..415M}, which is essentially the same in all star-based BH formation models. Notable exceptions include models where older populations of stars are responsible for present-day BHs \cite{2014MNRAS.442.2963K,2017MNRAS.471.4702B}, as well as predictions which make use of large-scale cosmological simulations \cite{2017MNRAS.472.2422M,2018MNRAS.480.2704L}.

BH spin magnitudes can be very powerful observables for constraining the physics of individual massive stars, but provide a less effective way to distinguish between stellar-based compact-binary formation channels. Spin magnitudes are expected to be set by stellar collapse dynamics \cite{2011ApJ...730...70O}, and should therefore be similar for BHs formed either in galactic fields or dynamically. A possible handle could be provided by dependence on the star's metallicity, which is expected to impact processes like angular momentum transport and mass loss.  Again, these observables might turn out to be particularly useful to constrain specific mechanisms, such as scenarios where previous mergers, rather than stellar collapse, are responsible for forming the merging BHs \cite{2017PhRvD..95l4046G,2017ApJ...840L..24F,2018PhRvD..97l3003K}.

Binary eccentricities may also provide information on some specific models \cite{2012ApJ...757...27A,2018PhRvD..97j3014S}. Eccentricities from the most favored scenarios are expected to be too low in the LIGO/Virgo band  to provide stringent constraints \cite{2017CQGra..34j4002A}, {although some scenarios predict events with high eccentricity \cite{2018PhRvD..97j3014S,2018PhRvL.120o1101R}}. To this end, LISA observations at low frequencies (when binaries are not yet fully circularized) may turn out to be crucial  \cite{2016PhRvD..94f4020N,2017MNRAS.465.4375N,2016ApJ...830L..18B,2018arXiv180406519S}.

The most promising observables to shine light on BH-binary formation are the spin directions. Spins of BHs formed
following dynamical encounters are expected to be isotropically distributed {(but see \cite{2017ApJ...846L..11L,2018MNRAS.480L..58A}).}
This is because the BH-binary evolution is set by the astrophysical environment, whose coupling to the BH spins is known to be negligible \cite{2017PhRvD..95f4014C}. It is worth pointing out, however, that some angular momentum from the cloud that formed the cluster could be transferred to the stellar spins, thus introducing correlations between their directions \cite{2017NatAs...1E..64C}. Even if present, these correlations are expected to be largely washed out by the many dynamical encounters leading to the formation of the GW sources. %

Conversely, the spin directions of BH binaries formed in isolation are greatly influenced by the evolutionary paths of their stellar progenitors. %
The two stars will form a binary BH without prominent interactions with other bodies, thus ``carrying memory'' of some of the physical mechanisms  occurring during their history.
Even in the simplest models where stellar spins are initially aligned to the binary's orbital angular momentum, misalignments are expected to be introduced by recoil velocities imparted to the BHs at birth. These ``supernova kicks'' tilt the orbital plane, thus introducing some misalignment between the orbital angular momentum and the spin directions \cite{2000ApJ...541..319K}. Tidal interactions
can also influence the spin directions, generically acting towards realigning spins with the orbital angular momentum \cite{1981A&A....99..126H}. After the BH binary is formed, spin directions are further modified by post-Newtonian (PN) spin-orbit and spin-spin couplings during the long inspiral phase before the binary becomes detectable by LIGO and Virgo \cite{1994PhRvD..49.6274A}. PN effects tend to separate different subpopulations, hence greatly improving model distinguishability \cite{2013PhRvD..87j4028G}. %
The effectiveness of BH spin tilts at constraining formation channels was already explored in previous work through astrophysical models \cite{2010CQGra..27k4007M,2013PhRvD..87j4028G,2016ApJ...832L...2R,
2018MNRAS.480L..58A,2017arXiv170607053B}, simulated LIGO/Virgo data \cite{2014PhRvD..89l4025G,2014PhRvL.112y1101V,2016PhRvD..93d4071T,2017MNRAS.471.2801S,2017CQGra..34cLT01V,2017PhRvD..96b3012T} and actual GW observations \cite{2017PhRvL.119a1101O,2017Natur.548..426F,2018PhRvD..97d3014W,2018ApJ...854L...9F,2018arXiv180506442W}.

This paper presents a comprehensive study of the expected spin direction distributions of BH binaries formed from isolated pairs of stars.
Using the \textsc{StarTrack} \cite{2008ApJS..174..223B}  and  \textsc{precession} \cite{2016PhRvD..93l4066G} numerical codes, we combine for the first time state-of-the-art evolutions of binary stars to accurate PN spin tracking and coherently model spin evolution from formation to detection (Sec.~\ref{methods}). We present forecasts for both approximate one-spin dynamics through the effective-spin parameter (which is easier to measure; Sec.~\ref{results1}) and genuine two-spin effects (which encode more information; Sec.~\ref{results2}). We then illustrate predictions of our models in terms of the spin morphologies identified in \cite{2015PhRvL.114h1103K,2015PhRvD..92f4016G} (Sec.~\ref{results3}). We conclude with   prospects for constraining these mechanisms with current and future GW detectors (Sec.~\ref{conclusions}). Unless otherwise noted, we use geometrical units ($G=c=1$).

Our database is publicly available at \href{https://github.com/dgerosa/spops}{github.com/dgerosa/spops} \cite{spops}, where we also provide a convenient python module (called \textsc{spops}) to facilitate its exploration.

\section{Methods: stellar and black-hole evolution}
\label{methods}

We perform binary-star evolutions using the collection of semianalytic prescriptions implemented in the \mbox{\textsc{StarTrack}} code \cite{2002ApJ...572..407B,2008ApJS..174..223B,2012ApJ...759...52D,2012ApJ...749...91F,2013ApJ...779...72D,2015ApJ...806..263D,2016A&A...594A..97B,2016ApJ...819..108B}. Each evolution results in a BH binary characterized by masses $m_i$ (with $i=1,2$; or alternatively $q=m_2/m_1\leq1$ and $M=m_1+m_2$), spin magnitudes $|\mathbf{S_i}| = \chi_i m_i^2$ (with $0\leq \chi_i\leq 1$), directions $\hat{\mathbf{S_i}}$, and merger-rate weight (see Sec.~\ref{detectability} below). The direction of each spin is described by a polar angles $\theta_i$ (relative to the direction of the orbital angular momentum) and by an azimuthal angle in the orbital plane, $\phi_i$.

Our suite of models is described below. Each model has a single free parameter $\sigma$ (setting the magnitude of the kick velocities) and three flags corresponding to our assumptions on spin magnitudes, tidal interactions and the sensitivity of GW detectors. With the exception of BH kicks and spins, all other assumptions are the same as in model M10 of \cite{2016A&A...594A..97B}. We refer the reader to that paper for a more comprehensive description of our population-synthesis simulations.

\subsection{Spin magnitudes}

As for the BH spin magnitudes, we implement three different models:
\begin{itemize}

\item[A1)] ``\emph{uniform}'': We assume the dimensionless BH Kerr parameters $\chi_i$ to be uniformly distributed in $[0,1]$, independently of the other binary parameters.

\begin{figure}
\includegraphics[width=\columnwidth]{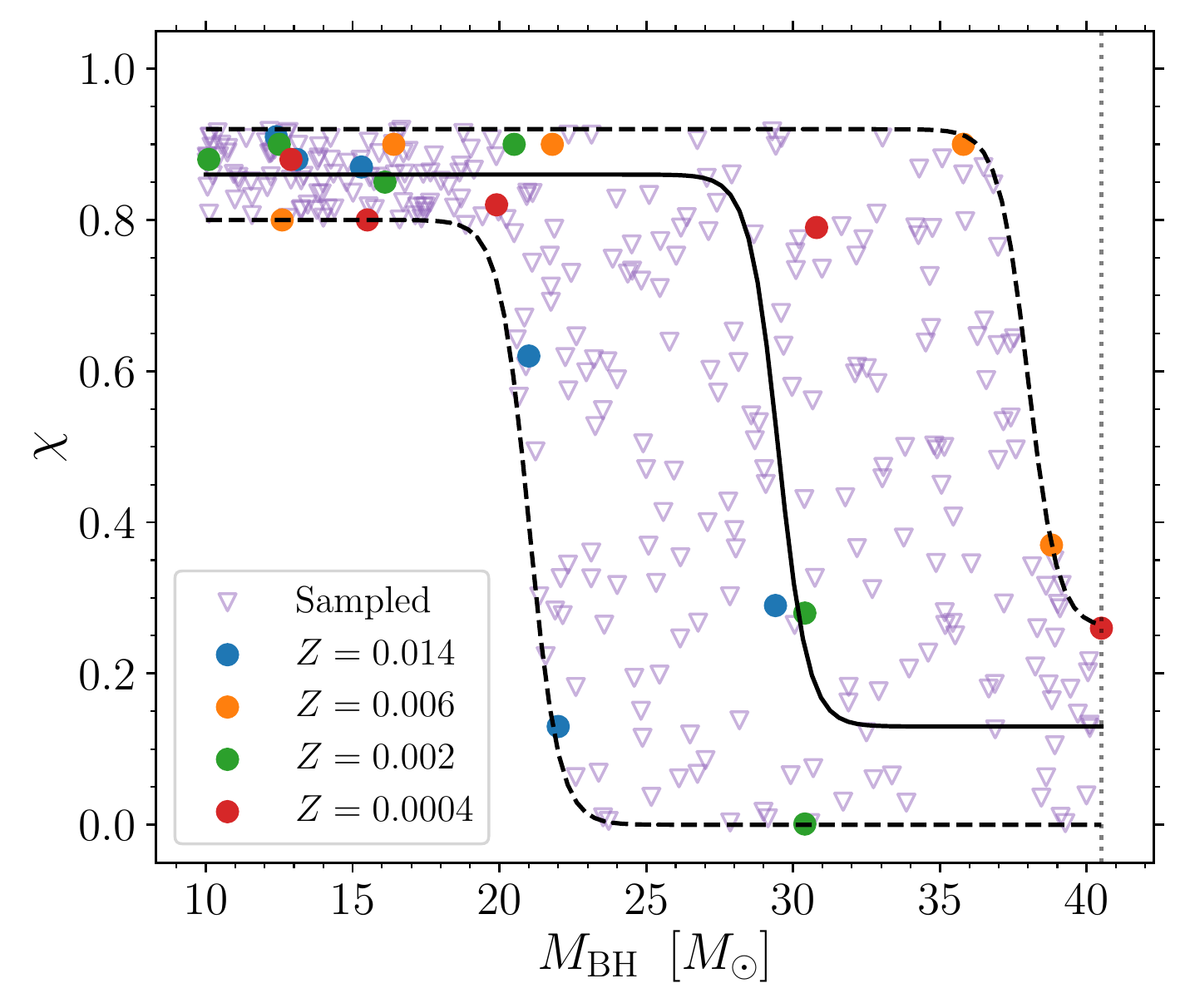}
\caption{Model \emph{collapse} for the BH spin magnitude as a function of the BH mass. In this model, heavier (lighter) collapsing stars preferentially form BHs with smaller (larger) spins. Filled circles shows data points from the simulations reported by \cite{2017arXiv170607053B} at various metallicities $Z$, while empty triangles show our resampled distribution. Dashed (solid) lines illustrate our construction procedure (see text). The hard cutoff at $M_{\rm BH}=40.5 M_\odot$ (dotted line) is due to pulsational pair-instability supernovae as implemented in \textsc{StarTrack} \cite{2016A&A...594A..97B}.}
\label{fitspin}
\end{figure}

\item[A2)] ``\emph{collapse}'': Simulations of stellar collapse show that stars with large (low) mass tend to form slowly (highly) rotating BHs \cite{2017arXiv170607053B,2018A&A...616A..28Q}. %
This feature introduces a specific correlation between masses and spins, with potentially critical impact on the predicted GW sources. Here we implement a very simple prescription to qualitatively capture this effect, leaving more robust explorations to future work. We %
use evolutionary simulations of stars with specific angular momentum transport 
 from \cite{2012A&A...544L...4E,2013A&A...558A.103G} as reported in Table~3  of \cite{2017arXiv170607053B}, together with the approximate expression 
\begin{equation}
M_{\rm BH} = \min(M_{\rm CO} + 3 M_\odot, 40.5 M_\odot)
\end{equation}
obtained from Fig.~1 of \cite{2017arXiv170607053B}, {where $M_{\rm CO}$ is the carbon-oxygen core mass.} Since there are not enough data points to construct meaningful interpolants, 
we opt for the following heuristic approach.  At low (large) masses, spins appear to be centered about $\chi\!\sim\!0.8$ ($0.13$) with a scatter of $\sim\!0.06$ ($0.13$). The turnover between the two regimes is at $M_{\rm BH}\!\sim\!29.5 M_\odot$ with a scatter of about $\sim \! 8.5 M_\odot$. Our procedure is illustrated in Fig.~\ref{fitspin}. We first construct two curves {to bracket the uncertainties:}
\begin{equation}
\chi =  \frac{p_1 - p_2}{2}  \tanh\left(p_3 - \frac{M_{\rm BH}}{ M_\odot}\right) + \frac{p_1 + p_2}{2}
\end{equation}
where $p_i=0.86\pm 0.06, 0.13\pm0.13, 29.5 \pm 8.5$ (the upper/lower signs refer to the upper/lower limits in Fig.~\ref{fitspin}).
Spins are then generated {by drawing random samples uniformly} in the region in between the two curves. We argue that this model captures some of the key features found in  \cite{2017arXiv170607053B}, namely that larger BH masses tend to correlate with  smaller spins, while at the same time reflecting the large uncertainties of those results.\footnote{While this work was being completed, a similar approximation was proposed in \cite{2018arXiv180601285A}. {Other parametrized spin models have been proposed in, e.g., \cite{2018arXiv180506442W,2017PhRvD..96b3012T}.}} 

\item[A3)] ``\emph{max}'': In order to maximize the effects of spin- precession dynamics and highlight some trends, we also run a set of models where all BHs are maximally spinning ($\chi_1=\chi_2=1$).

\end{itemize}

We note that the first two models implement rather conservative assumptions regarding the expected spin-precession
dynamics (cf. \cite{2013PhRvD..87j4028G,2017MNRAS.471.2801S}, where only very high spins are considered).
Our approach complements that of \cite{2017arXiv170607053B}, where a specific model [their Eq.~(3)] was assumed for the
 BH spin magnitude. %
More work is needed to fully include the impact of the metallicity on the expected BH spin magnitudes, which is here neglected.

\subsection{Spin directions}

We assume stellar spins to be initially aligned to the orbital angular momentum of the binary
($\theta_1=\theta_2=0$). {This same assumption} is made by most, if not all, population-synthesis models (but
see e.g. \cite{2012IAUS..282..397A,2017arXiv170600369P}). As the first star collapses and forms a BH, the resulting  kick tilts the orbital
plane \cite{2000ApJ...541..319K,2002MNRAS.329..897H} and introduces a spin-orbit misalignment ($\theta_1=\theta_2\neq
0$). It is worth pointing out that spin misalignments are induced by asymmetric mass and neutrino emission during core collapse (``natal kicks''), while symmetric mass loss only impacts the binary's center of mass (``Blaauw kicks''  \cite{1961BAN....15..265B}). 
Supernova kicks are drawn from a  Maxwellian distribution with 1D dispersion $\sigma$, independently of the mass of the system. We generate 7 different models\footnote{For future reference, these models were numbered M00, M18, M17, M16, M15, M14 and M13.} %
 at $\sigma=0,25,50,70,130,200$ and 265 km/s. The largest value $\sigma=265$ km/s corresponds to the observational constraints from pulsar proper-motion measurements \cite{2005MNRAS.360..974H}. We adopt this approach because it constitutes a simple and well-defined one-parameter family of models to illustrate the main trends of the BH spin alignment distributions. More elaborate (and perhaps more physical) prescriptions where, e.g., kicks are suppressed by fallback material \cite{1999ApJ...522..413F} can be constructed by appropriate mass-dependent mixture of our distributions \cite{2018PhRvD..97d3014W}.

After the first kick, tidal interactions may realign one of the spins. In between the two supernova explosions, the system is formed by a BH and a (perhaps evolved) star \cite{2016MNRAS.462..844K}. Since tidal interactions scale with the cube of the size of the object, tides raised on  the star by the BH are much more effective than tides raised on the BH by the star.

In the spirit of introducing only minimal assumptions, we implement three prescriptions  \cite{2013PhRvD..87j4028G,2017arXiv170607053B} and postpone a more careful treatment of tidal spin alignment to future work \cite{tidesprep}.

\begin{itemize}
\item[B1)] ``\emph{alltides}'': Tidal interactions align all stellar spins in between the two explosions. This corresponds to setting either $\theta_1\neq 0$, $\theta_2=0$ or $\theta_2\neq 0$, $\theta_1=0$, depending of which star explodes first.
\item[B2)] ``\emph{notides}'': None of the spins is realigned.
\item[B3)] ``\emph{time}'': We attempt a physical model for tidal interactions following \cite{2016MNRAS.462..844K,2018MNRAS.473.4174Z}. In particular they estimate the tidal alignment time  to be
\begin{align}
t_\tau &= 4 \times 10^{4}\;\; \left( \frac{M_{\rm S}}{M_{\rm BH}}\right)^2
\left( \frac{2 M_{\rm S}}{M_{\rm S}+ M_{\rm BH}}\right)^{5/6}
\notag \\
&\times \left( \frac{r}{R_\odot}\right)^{17/2}
\left( \frac{M_s}{M_\odot}\right)^{-51/8} {\rm yr}\;,
\label{ttau}
\end{align}
where $M_{\rm S}$ and $M_{\rm BH}$ are the masses of the star and the BH, respectively, and $r$ is the binary separation. We compute Eq.~(\ref{ttau}) from the \textsc{StarTrack} data before the second supernova, and compare the result with both the time between the two explosions $t_{\rm SN}$ and the typical lifetime of a Wolf-Rayet star $t_{\rm WR} =3 \cdot 10^5$ yr \cite{2016MNRAS.462..844K,2018MNRAS.473.4174Z}. The star's spin is realigned if 
$
t_{\tau}<\min(t_{\rm SN}, t_{\rm WR}).
$
\end{itemize}

The azimuthal angles $\phi_i$ may also evolve in between the two explosions because of relativistic spin precession. We compare the time between the two explosions $t_{\rm SN}$ to the leading-order precession timescale \cite{2015PhRvD..92f4016G}
\begin{equation}
t_{\rm pre} = M \frac{4 \pi}{3} \frac{1 + q}{1 - q}  \left(\frac{r}{M}\right)^{5/2}\,.
\end{equation}
{Binaries have $\phi_1=\phi_2=0$ after the first SN; these angles are updated only if there is enough time for the spins to precess before the second explosion. Therefore we set $\phi_i=0$ if $t_{\rm SN}<t_{\rm pre}$, or draw $\phi_i$ randomly if $t_{\rm SN}>t_{\rm pre}$.}

It is worth noting that tidal interactions (here considered only regarding the spin directions) are expected to affect the spin magnitude of the second-born BH as well \cite{2016MNRAS.462..844K,2018MNRAS.473.4174Z,2018A&A...616A..28Q}. Tidally locked stars are going to be both aligned and spun up. This behavior is partially captured by the combination of the \emph{collapse} and \emph{time} models, as lower mass stars are both assigned high spins and lower $t_{\rm \tau}$.  A more systematic study of the effect of tides in BH binaries formation pathways is under development \cite{tidesprep}. The potential impact of mass transfer on both spin magnitudes and directions is also an avenue of future improvement.

At the second explosion, another supernova kick will further tilt  the orbital plane. This finally results in the angles $\theta_1, \theta_2$ and $\Delta\Phi=\phi_2-\phi_1$ at BH-binary formation.
Each system now needs to be evolved from the separation where it forms down the LIGO/Virgo band ($f_{\rm GW} = 20$ Hz). We use the numerical code \textsc{precession} \cite{2016PhRvD..93l4066G}, which implements multi-timescale methods to efficiently evolve BH systems over their very long inspirals  before merger \cite{2015PhRvD..92f4016G,2015PhRvL.114h1103K}. This is a crucial improvement over previous work \cite{2013PhRvD..87j4028G,2017MNRAS.471.2801S} (but see \cite{2016ApJ...832L...2R,2018PhRvD..97d3014W}, which also use the same code), since it was shown that integrations from separations  as large as $\sim\!10^6 M$ may be necessary to fully capture spin-precession effects \cite{2015PhRvD..92f4016G}.

\subsection{Detectability}
\label{detectability}

For each set of assumptions, our procedure generates a sample of weighted BH binaries \cite{2013ApJ...779...72D,2015ApJ...806..263D,2016ApJ...819..108B}. These weights correspond to the merger rate contribution of each evolutionary track, taking into account redshift- and metallicity-dependent star formation history  \cite{2013ApJ...779...72D}, as well as  the antenna pattern of the GW interferometer \cite{2015ApJ...806..263D}.

Contrary to previous {\sc StarTrack} studies, we now take into account spin corrections in the calculation of the merger
weight. As illustrated below, this is a crucial point to faithfully predict spin distributions which, if neglected,
could lead to sizable biases \cite{2018arXiv180503046N,2013PhRvD..87b4035B,2015ApJ...806..263D,2010PhRvD..82j4006O,2009PhRvD..80l4026R}. 
We generate GW signals using the \textsc{IMRPhenomPv2} \cite{2014PhRvL.113o1101H} waveform model as implemented in the \textsc{pyCBC} pipeline \cite{2016CQGra..33u5004U}. We compute signal-to-noise ratios (SNRs) using three different noise curves:
\begin{itemize}
\item[C1)] \emph{LIGO:} the expected sensitivity for Advanced LIGO in its design configuration \cite{2018LRR....21....3A};
\item[C2)]  \emph{Voyager:} a planned upgrade designed to maximize the science return within the current LIGO facilities \cite{Voyagercurve};
\item[C3)]  \emph{Cosmic Explorer:} a proposed third-generation detector in 40-km scale facilities  \cite{2017CQGra..34d4001A}.
\end{itemize}
For simplicity we consider single detectors with a SNR threshold of 8, which is a reasonable approximation to mimic realistic data analysis procedures \cite{2010CQGra..27q3001A}.  For reference, an optimally located and oriented, equal-mass, nonspinning BH binary with source-frame total mass of $60M_\odot$ will have an SNR larger than 8 at redshifts $z\lesssim 1.2$ for Advanced LIGO, $z\lesssim 9.1$ for Voyager and $z\lesssim 36.4$ for Cosmic Explorer. Detection rates $r$ (in units of ${\rm yr}^{-1}$)  are then computed as detailed in \cite{2016ApJ...819..108B} (see also \cite{1993PhRvD..47.2198F,1996PhRvD..53.2878F,2015ApJ...806..263D}) using the public code \textsc{gwdet} \cite{gwdet}.

\begin{figure}
\includegraphics[width=\columnwidth]{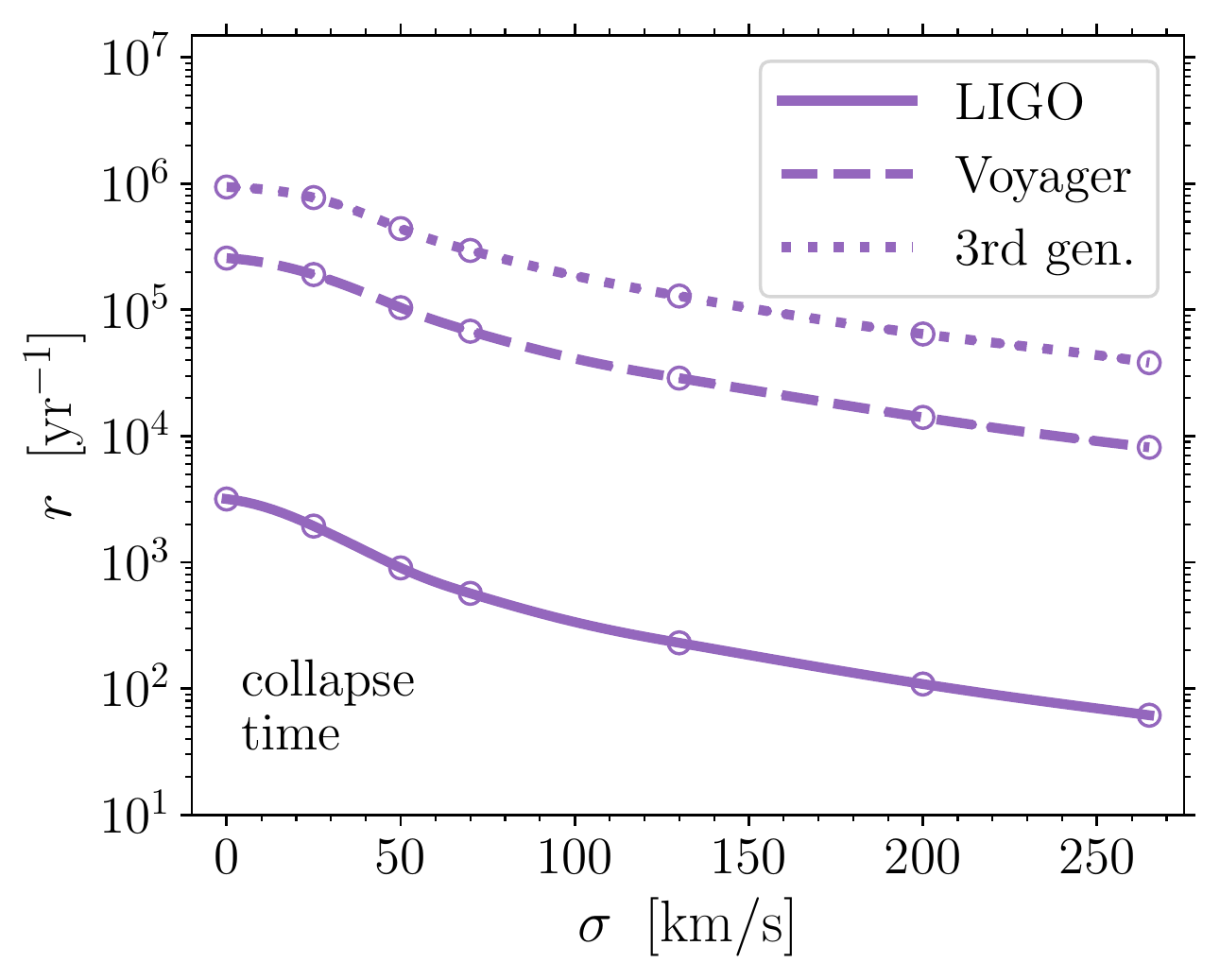}
\caption{Detection rates for LIGO (solid line), Voyager (dashed line) and Cosmic Explorer (dotted line) for models with different kick speed parameters $\sigma$, assuming the \emph{time} tidal model and the \emph{collapse} spin model. All other models give qualitatively similar curves; the \emph{max} spin model yields marginally higher rates for LIGO (cf. Table~\ref{ratestable}). {For a discussion of uncertainties in  detection rates see e.g.~\cite{2013ApJ...779...72D,2015ApJ...806..263D}.}}
\label{overallrates}
\end{figure}

\begin{table*}
\begin{tabular}{@{\hskip 0.1in}c@{\hskip 0.1in}@{\hskip 0.1in}c@{\hskip 0.1in}@{\hskip 0.1in}c@{\hskip 0.2in}|@{\hskip 0.2in}c@{\hskip 0.1in}@{\hskip 0.1in}c@{\hskip 0.1in}@{\hskip 0.1in}c@{\hskip 0.1in}@{\hskip 0.1in}c@{\hskip 0.1in}@{\hskip 0.1in}c@{\hskip 0.1in}@{\hskip 0.1in}c@{\hskip 0.1in}@{\hskip 0.1in}c@{\hskip 0.1in}}
{\bf Detector} & {\bf Spins} & {\bf Tides} & \multicolumn{7 }{c}{\bf Natal kick $\boldsymbol\sigma$} \\
 & & &   $0$ km/s & $25$ km/s & $50$ km/s & $70$ km/s & $130$ km/s & $200$ km/s & $265$ km/s\\
\hline
\hline
LIGO & collapse & time&  $ 3.2\!\times\! 10^3$ &$ 1.9\!\times\! 10^3$ &$ 9.0\!\times\! 10^2$ &$ 5.7\!\times\! 10^2$ &$ 2.3\!\times\! 10^2$ &$ 1.1\!\times\! 10^2$ &$ 6.1\!\times\! 10^1$  \\
LIGO & collapse & alltides&  $ 3.2\!\times\! 10^3$ &$ 2.0\!\times\! 10^3$ &$ 9.3\!\times\! 10^2$ &$ 5.9\!\times\! 10^2$ &$ 2.4\!\times\! 10^2$ &$ 1.1\!\times\! 10^2$ &$ 6.3\!\times\! 10^1$  \\
LIGO & collapse & notides&  $ 3.2\!\times\! 10^3$ &$ 1.9\!\times\! 10^3$ &$ 8.9\!\times\! 10^2$ &$ 5.6\!\times\! 10^2$ &$ 2.2\!\times\! 10^2$ &$ 1.1\!\times\! 10^2$ &$ 6.0\!\times\! 10^1$  \\
LIGO & uniform & time&  $ 3.2\!\times\! 10^3$ &$ 1.9\!\times\! 10^3$ &$ 8.8\!\times\! 10^2$ &$ 5.6\!\times\! 10^2$ &$ 2.3\!\times\! 10^2$ &$ 1.2\!\times\! 10^2$ &$ 6.3\!\times\! 10^1$  \\
LIGO & uniform & alltides&  $ 3.2\!\times\! 10^3$ &$ 1.9\!\times\! 10^3$ &$ 9.2\!\times\! 10^2$ &$ 5.9\!\times\! 10^2$ &$ 2.4\!\times\! 10^2$ &$ 1.2\!\times\! 10^2$ &$ 6.4\!\times\! 10^1$  \\
LIGO & uniform & notides&  $ 3.2\!\times\! 10^3$ &$ 1.9\!\times\! 10^3$ &$ 8.7\!\times\! 10^2$ &$ 5.5\!\times\! 10^2$ &$ 2.3\!\times\! 10^2$ &$ 1.1\!\times\! 10^2$ &$ 6.1\!\times\! 10^1$  \\
LIGO & max & time&  $ 5.1\!\times\! 10^3$ &$ 3.0\!\times\! 10^3$ &$ 1.3\!\times\! 10^3$ &$ 8.1\!\times\! 10^2$ &$ 3.3\!\times\! 10^2$ &$ 1.7\!\times\! 10^2$ &$ 8.8\!\times\! 10^1$  \\
LIGO & max & alltides&  $ 5.1\!\times\! 10^3$ &$ 3.0\!\times\! 10^3$ &$ 1.4\!\times\! 10^3$ &$ 8.6\!\times\! 10^2$ &$ 3.5\!\times\! 10^2$ &$ 1.7\!\times\! 10^2$ &$ 9.2\!\times\! 10^1$  \\
LIGO & max & notides&  $ 5.1\!\times\! 10^3$ &$ 2.9\!\times\! 10^3$ &$ 1.3\!\times\! 10^3$ &$ 7.8\!\times\! 10^2$ &$ 3.2\!\times\! 10^2$ &$ 1.6\!\times\! 10^2$ &$ 8.4\!\times\! 10^1$  \\
Voyager & collapse & time&  $ 2.6\!\times\! 10^5$ &$ 1.9\!\times\! 10^5$ &$ 1.0\!\times\! 10^5$ &$ 6.8\!\times\! 10^4$ &$ 2.9\!\times\! 10^4$ &$ 1.4\!\times\! 10^4$ &$ 8.1\!\times\! 10^3$  \\
Voyager & collapse & alltides&  $ 2.6\!\times\! 10^5$ &$ 1.9\!\times\! 10^5$ &$ 1.0\!\times\! 10^5$ &$ 6.9\!\times\! 10^4$ &$ 2.9\!\times\! 10^4$ &$ 1.4\!\times\! 10^4$ &$ 8.2\!\times\! 10^3$  \\
Voyager & collapse & notides&  $ 2.6\!\times\! 10^5$ &$ 1.9\!\times\! 10^5$ &$ 1.0\!\times\! 10^5$ &$ 6.5\!\times\! 10^4$ &$ 2.7\!\times\! 10^4$ &$ 1.3\!\times\! 10^4$ &$ 7.7\!\times\! 10^3$  \\
Voyager & uniform & time&  $ 2.3\!\times\! 10^5$ &$ 1.7\!\times\! 10^5$ &$ 9.6\!\times\! 10^4$ &$ 6.3\!\times\! 10^4$ &$ 2.8\!\times\! 10^4$ &$ 1.4\!\times\! 10^4$ &$ 8.0\!\times\! 10^3$  \\
Voyager & uniform & alltides&  $ 2.3\!\times\! 10^5$ &$ 1.7\!\times\! 10^5$ &$ 9.6\!\times\! 10^4$ &$ 6.3\!\times\! 10^4$ &$ 2.8\!\times\! 10^4$ &$ 1.4\!\times\! 10^4$ &$ 8.1\!\times\! 10^3$  \\
Voyager & uniform & notides&  $ 2.3\!\times\! 10^5$ &$ 1.7\!\times\! 10^5$ &$ 9.4\!\times\! 10^4$ &$ 6.1\!\times\! 10^4$ &$ 2.6\!\times\! 10^4$ &$ 1.3\!\times\! 10^4$ &$ 7.7\!\times\! 10^3$  \\
Voyager & max & time&  $ 3.0\!\times\! 10^5$ &$ 2.2\!\times\! 10^5$ &$ 1.2\!\times\! 10^5$ &$ 7.8\!\times\! 10^4$ &$ 3.3\!\times\! 10^4$ &$ 1.6\!\times\! 10^4$ &$ 9.6\!\times\! 10^3$  \\
Voyager & max & alltides&  $ 3.0\!\times\! 10^5$ &$ 2.2\!\times\! 10^5$ &$ 1.2\!\times\! 10^5$ &$ 8.0\!\times\! 10^4$ &$ 3.4\!\times\! 10^4$ &$ 1.7\!\times\! 10^4$ &$ 9.8\!\times\! 10^3$  \\
Voyager & max & notides&  $ 3.0\!\times\! 10^5$ &$ 2.2\!\times\! 10^5$ &$ 1.2\!\times\! 10^5$ &$ 7.5\!\times\! 10^4$ &$ 3.1\!\times\! 10^4$ &$ 1.5\!\times\! 10^4$ &$ 8.9\!\times\! 10^3$  \\
3$^{\rm rd}$ gen. & collapse & time&  $ 9.4\!\times\! 10^5$ &$ 7.7\!\times\! 10^5$ &$ 4.4\!\times\! 10^5$ &$ 3.0\!\times\! 10^5$ &$ 1.3\!\times\! 10^5$ &$ 6.4\!\times\! 10^4$ &$ 3.8\!\times\! 10^4$  \\
3$^{\rm rd}$ gen. & collapse & alltides&  $ 9.4\!\times\! 10^5$ &$ 7.7\!\times\! 10^5$ &$ 4.4\!\times\! 10^5$ &$ 3.0\!\times\! 10^5$ &$ 1.3\!\times\! 10^5$ &$ 6.4\!\times\! 10^4$ &$ 3.8\!\times\! 10^4$  \\
3$^{\rm rd}$ gen. & collapse & notides&  $ 9.4\!\times\! 10^5$ &$ 7.7\!\times\! 10^5$ &$ 4.4\!\times\! 10^5$ &$ 2.9\!\times\! 10^5$ &$ 1.3\!\times\! 10^5$ &$ 6.4\!\times\! 10^4$ &$ 3.8\!\times\! 10^4$  \\
3$^{\rm rd}$ gen. & uniform & time&  $ 9.4\!\times\! 10^5$ &$ 7.7\!\times\! 10^5$ &$ 4.4\!\times\! 10^5$ &$ 2.9\!\times\! 10^5$ &$ 1.3\!\times\! 10^5$ &$ 6.4\!\times\! 10^4$ &$ 3.8\!\times\! 10^4$  \\
3$^{\rm rd}$ gen. & uniform & alltides&  $ 9.4\!\times\! 10^5$ &$ 7.7\!\times\! 10^5$ &$ 4.4\!\times\! 10^5$ &$ 2.9\!\times\! 10^5$ &$ 1.3\!\times\! 10^5$ &$ 6.4\!\times\! 10^4$ &$ 3.8\!\times\! 10^4$  \\
3$^{\rm rd}$ gen. & uniform & notides&  $ 9.4\!\times\! 10^5$ &$ 7.7\!\times\! 10^5$ &$ 4.4\!\times\! 10^5$ &$ 2.9\!\times\! 10^5$ &$ 1.3\!\times\! 10^5$ &$ 6.4\!\times\! 10^4$ &$ 3.8\!\times\! 10^4$  \\
3$^{\rm rd}$ gen. & max & time&  $ 9.4\!\times\! 10^5$ &$ 7.8\!\times\! 10^5$ &$ 4.4\!\times\! 10^5$ &$ 3.0\!\times\! 10^5$ &$ 1.3\!\times\! 10^5$ &$ 6.4\!\times\! 10^4$ &$ 3.8\!\times\! 10^4$  \\
3$^{\rm rd}$ gen. & max & alltides&  $ 9.4\!\times\! 10^5$ &$ 7.8\!\times\! 10^5$ &$ 4.4\!\times\! 10^5$ &$ 3.0\!\times\! 10^5$ &$ 1.3\!\times\! 10^5$ &$ 6.4\!\times\! 10^4$ &$ 3.8\!\times\! 10^4$  \\
3$^{\rm rd}$ gen. & max & notides&  $ 9.4\!\times\! 10^5$ &$ 7.8\!\times\! 10^5$ &$ 4.4\!\times\! 10^5$ &$ 2.9\!\times\! 10^5$ &$ 1.3\!\times\! 10^5$ &$ 6.4\!\times\! 10^4$ &$ 3.8\!\times\! 10^4$  \\
\end{tabular}
\caption{Detection rates in units of yr$^{-1}$ for all our simulations. Results for LIGO assume the expected design sensitivity of the instrument \cite{2016CQGra..33u5004U}. Voyager is a planned instrumental upgrade to be located in the current LIGO facilities \cite{Voyagercurve}. Here we use Cosmic Explorer \cite{2017CQGra..34d4001A} as an illustrative example of what would be possible with third-generation detectors. {For a discussion of uncertainties in  detection rates see e.g.~\cite{2013ApJ...779...72D,2015ApJ...806..263D}.}}
\label{ratestable}
\end{table*}

The detection rate $r$ is a steep function of the natal kick velocity  \cite{2002ApJ...572..407B}. For large values of $\sigma$, more and more stellar progenitor binaries are unbound by natal kicks and fail to form GW sources.  This is shown in Fig.~\ref{overallrates} for a subset of our models; we find that $r$ drops by about a factor $\sim\!20$ between $\sigma=25$ km/s and $\sigma=265$ km/s. Third-generation detectors increase the expected detection rates by a factor of $\sim 300$ compared to LIGO at design sensitivity \cite{2017CQGra..34d4001A}.

Detection rates for each of our model variations are reported in Table~\ref{ratestable}. For LIGO, the \emph{max} spin models predicts higher rates (by about $\sim 50\%$, i.e a factor 1.5) when compared to models with \emph{uniform} spin distributions.  This behavior is due to the orbital hangup, a well-know effect in BH binary dynamics which causes binaries with aligned (anti-aligned) spins to have a larger (smaller) horizon distance \cite{2006PhRvD..74d1501C,2015CQGra..32j5009S}.   For distributions with spins mostly aligned like ours, binaries with large spin magnitudes are therefore easier to detect.   This result refines the rough estimate of \cite{2015ApJ...806..263D}, where spins were estimated to increase rates by at most a factor of 3 (cf. also \cite{2009PhRvD..80l4026R})
Interestingly, this rate increase disappears for third-generation detectors: future instruments will detect virtually all stellar-mass BH mergers in the Universe, irrespectively of their spins. For the same reason, models with more misaligned spins (\emph{notides}) have marginally lower rates than models where all spins are realigned by tidal interactions (\emph{alltides}).

\subsection{Simplified pathway classifications}
\label{pathway}
{\sc StarTrack} provides full information on the various  processes and the stellar types involved during each phase of the binary-star evolution. %
For this paper, we found particularly illustrative to simplify the classification of the evolutionary pathways marking the formation of the heavier BH (label ``BH1''), the formation of the lighter BH (label ``BH2'') and the occurrence of common envelope phases (label ``CE''). All of our stellar evolutions can be classified into eight mutually exclusive channels
\begin{center}
\begin{tabular}{l@{\hskip 0.5in}l}
{\bf 1.}$\;\;\;$BH1 CE BH2 &
{\bf 5.}$\;\;\;$CE BH1 CE BH2\\
{\bf 2.}$\;\;\;$BH2 CE BH1&
{\bf 6.}$\;\;\;$CE BH2 CE BH1\\
{\bf 3.}$\;\;\;$CE BH1 BH2&
{\bf 7.}$\;\;\;$BH1 BH2\\
{\bf 4.}$\;\;\;$CE BH2 BH1 &
{\bf 8.}$\;\;\;$BH2 BH1
\end{tabular}
\end{center}

{The abbreviations in the name of each channel should be intended as a chronological description of the events. For instance,}
the vanilla field-binary formation channel corresponds to the first case, ``BH1 CE BH2'': the heavier star collapses first and forms the heavier BH, a common-envelope phase tightens the binary, and finally the companion star forms the lighter BH. The second channel, ``BH2 CE BH1'', corresponds to cases where the light BH formed first: such mass-ratio reversal is known to have potentially strong impact on the  spin distribution \cite{2010PhRvD..81h4054K,2013PhRvD..87j4028G}.

\begin{figure}
\includegraphics[width=\columnwidth]{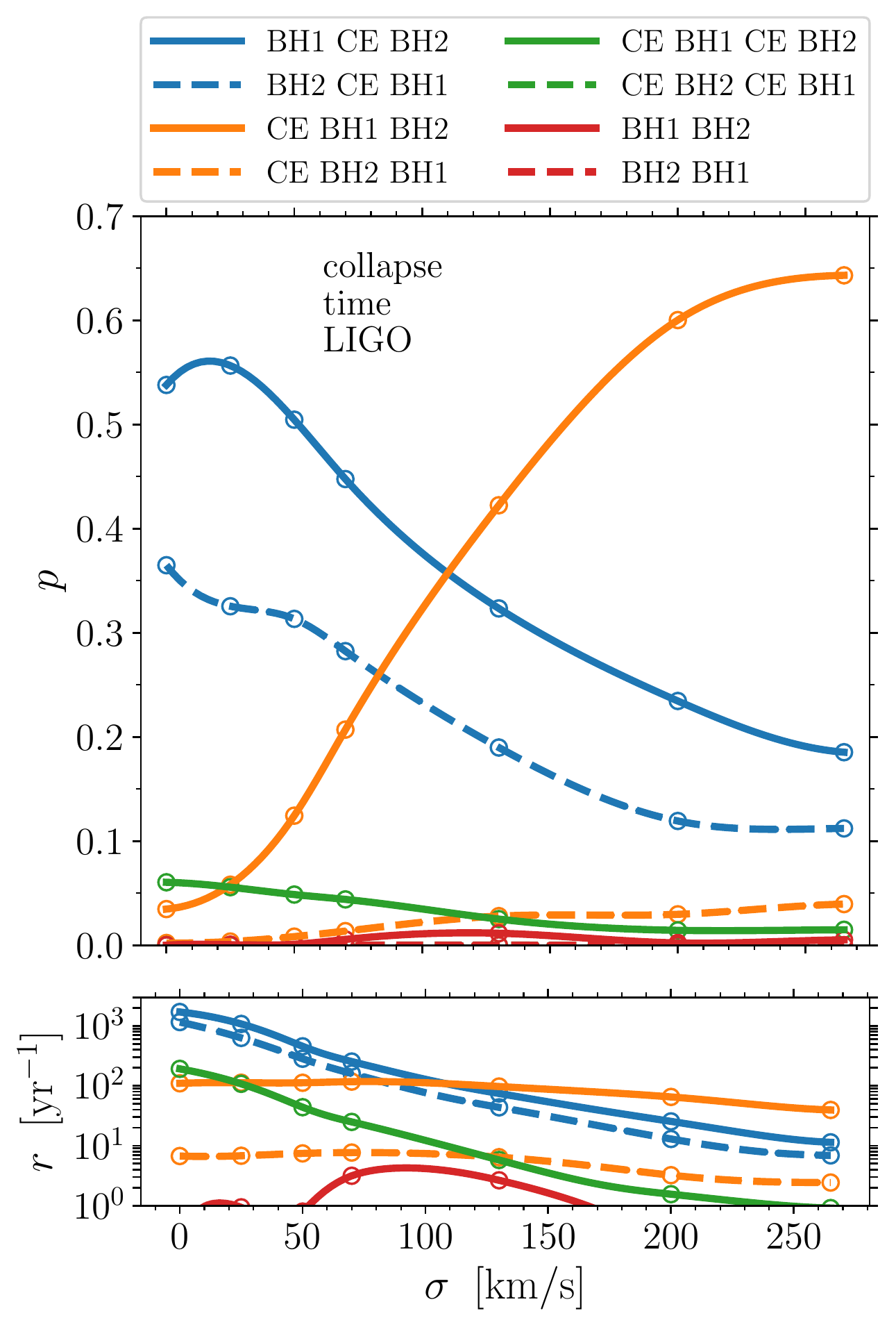}
\caption{Detection rates $r$ (bottom panel) and normalized rate fractions $p$ (top panels) of BH binaries formed via different channels as a function of the natal kicks $\sigma$. In this notation ``BH1'' and ``BH2'' stand for the formation of the heavier and lighter BH, respectively, while ``CE'' stand for the occurrence of a common-envelope phase (cf. Sec.~\ref{pathway}). Results are shown for the \emph{time} and \emph{collapse} spin model and weighted by LIGO detection rates. Results for Cosmic Explorer and other spin models are qualitatively similar.}
\label{ratechannels}
\end{figure}
The detection rates $r$ and their fraction in each channel  $p$  are shown in Fig.~\ref{ratechannels} as a function of the kick-velocity dispersion parameter $\sigma$~\cite{2005MNRAS.360..974H}.
For small kicks, most binaries follow the standard picture and evolve through a common-envelope phase between the two stellar collapses. At $\sigma=25 {\rm km/s}$ we have $p({\rm BH1\,CE\,BH2})+p({\rm BH2\,CE\,BH1})\sim 0.95$. Two thirds of these binaries follow the more standard pathway where the large BH is formed first,
while the rest undergo mass-ratio reversal.

If kicks are larger, the majority of binaries are found in the ``CE BH1 BH2'' channel. %
In this regime, systems are typically unbound by the first explosion (causing a drop in the rates), unless a common-envelope phase takes place before the explosion. Common envelope shrinks the orbital separation by orders of magnitude, thus dramatically increasing the chance of the binary surviving the first natal kick. In particular, in the extreme case $\sigma=265$ km/s we obtain  $p({\rm CE\,BH1\,BH2})\sim 0.65$.

Channels with zero or two common-envelope phases are always subdominant, and represent at most $p\lesssim 5\%$ of the population.

{Uncertainties in common-evenlope efficiency and more elaborate kick prescriptions are not explored in this paper (see e.g.~\cite{2012ApJ...759...52D,2018MNRAS.480.2011G,2018arXiv180608365T}) and might affect some of the results presented in Fig.~\ref{ratechannels}.}

\section{Results: effective spin}
\label{results1}

The spin parameter which is currently best measured in GWs \cite{2016PhRvX...6d1015A} is the effective spin \cite{2001PhRvD..64l4013D,2008PhRvD..78d4021R,2010PhRvD..81h4054K}
\begin{equation}
\chi_{\rm eff} = \frac{\chi_1 \cos\theta_1 + 	q \chi_2 \cos\theta_2}{1+q}
\label{chieff}
\end{equation}
(this is equivalent to $\xi$ in the notation of  \cite{2016PhRvD..93l4066G,2015PhRvD..92f4016G,2015PhRvL.114h1103K,2017CQGra..34f4004G,2017PhRvD..96b4007Z}). The effective spin is a constant of motion at 2PN order \cite{2008PhRvD..78d4021R,2010PhRvD..81h4054K}, and is therefore an excellent parameter to parametrize the dynamics  because it depends very weakly on the frequency/time at which it is measured by parameter-estimation algorithms.%

As evident from the definition (\ref{chieff}), measurements of $\chi_{\rm eff}$ are inevitably plagued by a degeneracy between the spin magnitudes and their directions: small (large) values of $\chi_{\rm eff}$ could be realized by either small (large) spin magnitudes or large (small) misalignment angles. For strategies to maximize the astrophysics that can be inferred from measurements of $\chi_{\rm eff}$ alone, see e.g.~\cite{2017Natur.548..426F,2018ApJ...854L...9F,2018arXiv180503046N,2018ApJ...854L...9F}.

\subsection{Marginalized distributions}
\label{margdistchieff}

\begin{table*}
\begin{tabular}{@{\hskip 0.1in}c@{\hskip 0.1in}@{\hskip 0.1in}c@{\hskip 0.1in}@{\hskip 0.1in}c@{\hskip 0.2in}|@{\hskip 0.2in}c@{\hskip 0.1in}@{\hskip 0.1in}c@{\hskip 0.1in}@{\hskip 0.1in}c@{\hskip 0.1in}@{\hskip 0.1in}c@{\hskip 0.1in}@{\hskip 0.1in}c@{\hskip 0.1in}@{\hskip 0.1in}c@{\hskip 0.1in}@{\hskip 0.1in}c@{\hskip 0.1in}}
{\bf Detector} & {\bf Spins} & {\bf Tides} & \multicolumn{7 }{c}{\bf Natal kick $\boldsymbol\sigma$} \\
 & & &   $0$ km/s & $25$ km/s & $50$ km/s & $70$ km/s & $130$ km/s & $200$ km/s & $265$ km/s\\
\hline
\hline
LIGO & collapse & time&  $ 0.79^{+0.10}_{-0.67} $ &$ 0.72^{+0.16}_{-0.61} $ &$ 0.67^{+0.21}_{-0.60} $ &$ 0.60^{+0.27}_{-0.62} $ &$ 0.49^{+0.37}_{-0.68} $ &$ 0.42^{+0.43}_{-0.61} $ &$ 0.38^{+0.46}_{-0.52} $  \\
LIGO & collapse & alltides&  $ 0.79^{+0.10}_{-0.66} $ &$ 0.73^{+0.15}_{-0.61} $ &$ 0.68^{+0.19}_{-0.58} $ &$ 0.63^{+0.24}_{-0.56} $ &$ 0.51^{+0.35}_{-0.48} $ &$ 0.42^{+0.43}_{-0.41} $ &$ 0.39^{+0.45}_{-0.45} $  \\
LIGO & collapse & notides&  $ 0.79^{+0.10}_{-0.67} $ &$ 0.71^{+0.17}_{-0.60} $ &$ 0.64^{+0.23}_{-0.60} $ &$ 0.57^{+0.30}_{-0.65} $ &$ 0.44^{+0.41}_{-0.78} $ &$ 0.41^{+0.43}_{-0.71} $ &$ 0.39^{+0.45}_{-0.66} $  \\
LIGO & uniform & time&  $ 0.54^{+0.33}_{-0.35} $ &$ 0.51^{+0.33}_{-0.36} $ &$ 0.46^{+0.36}_{-0.39} $ &$ 0.45^{+0.38}_{-0.49} $ &$ 0.41^{+0.40}_{-0.59} $ &$ 0.44^{+0.39}_{-0.63} $ &$ 0.36^{+0.41}_{-0.55} $  \\
LIGO & uniform & alltides&  $ 0.54^{+0.33}_{-0.35} $ &$ 0.52^{+0.33}_{-0.37} $ &$ 0.49^{+0.35}_{-0.37} $ &$ 0.47^{+0.35}_{-0.38} $ &$ 0.43^{+0.36}_{-0.42} $ &$ 0.43^{+0.39}_{-0.44} $ &$ 0.37^{+0.36}_{-0.45} $  \\
LIGO & uniform & notides&  $ 0.54^{+0.33}_{-0.35} $ &$ 0.50^{+0.33}_{-0.38} $ &$ 0.46^{+0.36}_{-0.41} $ &$ 0.42^{+0.39}_{-0.52} $ &$ 0.40^{+0.40}_{-0.66} $ &$ 0.42^{+0.37}_{-0.66} $ &$ 0.39^{+0.40}_{-0.63} $  \\
LIGO & max & time&  $ 1.00^{+0.00}_{0.00} $ &$ 0.99^{+0.01}_{-0.19} $ &$ 0.98^{+0.02}_{-0.45} $ &$ 0.97^{+0.03}_{-0.69} $ &$ 0.96^{+0.04}_{-0.88} $ &$ 0.94^{+0.06}_{-0.80} $ &$ 0.91^{+0.08}_{-0.82} $  \\
LIGO & max & alltides&  $ 1.00^{+0.00}_{0.00} $ &$ 0.99^{+0.01}_{-0.13} $ &$ 0.98^{+0.02}_{-0.36} $ &$ 0.97^{+0.03}_{-0.56} $ &$ 0.96^{+0.04}_{-0.79} $ &$ 0.94^{+0.05}_{-0.78} $ &$ 0.93^{+0.07}_{-0.81} $  \\
LIGO & max & notides&  $ 1.00^{+0.00}_{0.00} $ &$ 0.99^{+0.01}_{-0.21} $ &$ 0.97^{+0.03}_{-0.50} $ &$ 0.96^{+0.03}_{-0.71} $ &$ 0.96^{+0.04}_{-1.00} $ &$ 0.94^{+0.06}_{-0.82} $ &$ 0.90^{+0.10}_{-0.99} $  \\
Voyager & collapse & time&  $ 0.83^{+0.06}_{-0.69} $ &$ 0.80^{+0.08}_{-0.67} $ &$ 0.75^{+0.13}_{-0.65} $ &$ 0.69^{+0.18}_{-0.64} $ &$ 0.57^{+0.30}_{-0.60} $ &$ 0.48^{+0.37}_{-0.58} $ &$ 0.45^{+0.40}_{-0.52} $  \\
Voyager & collapse & alltides&  $ 0.83^{+0.06}_{-0.69} $ &$ 0.81^{+0.08}_{-0.67} $ &$ 0.76^{+0.12}_{-0.65} $ &$ 0.71^{+0.16}_{-0.63} $ &$ 0.59^{+0.28}_{-0.54} $ &$ 0.49^{+0.37}_{-0.47} $ &$ 0.46^{+0.39}_{-0.49} $  \\
Voyager & collapse & notides&  $ 0.83^{+0.06}_{-0.69} $ &$ 0.79^{+0.09}_{-0.67} $ &$ 0.72^{+0.16}_{-0.65} $ &$ 0.66^{+0.21}_{-0.68} $ &$ 0.54^{+0.33}_{-0.83} $ &$ 0.45^{+0.41}_{-0.82} $ &$ 0.44^{+0.41}_{-0.75} $  \\
Voyager & uniform & time&  $ 0.52^{+0.34}_{-0.35} $ &$ 0.50^{+0.33}_{-0.36} $ &$ 0.47^{+0.35}_{-0.37} $ &$ 0.45^{+0.36}_{-0.41} $ &$ 0.41^{+0.38}_{-0.50} $ &$ 0.41^{+0.37}_{-0.53} $ &$ 0.39^{+0.38}_{-0.52} $  \\
Voyager & uniform & alltides&  $ 0.52^{+0.34}_{-0.35} $ &$ 0.51^{+0.33}_{-0.36} $ &$ 0.48^{+0.34}_{-0.36} $ &$ 0.46^{+0.36}_{-0.37} $ &$ 0.42^{+0.38}_{-0.41} $ &$ 0.41^{+0.38}_{-0.44} $ &$ 0.39^{+0.37}_{-0.46} $  \\
Voyager & uniform & notides&  $ 0.52^{+0.33}_{-0.35} $ &$ 0.50^{+0.34}_{-0.37} $ &$ 0.45^{+0.36}_{-0.40} $ &$ 0.43^{+0.37}_{-0.50} $ &$ 0.40^{+0.39}_{-0.69} $ &$ 0.39^{+0.40}_{-0.72} $ &$ 0.36^{+0.38}_{-0.67} $  \\
Voyager & max & time&  $ 1.00^{+0.00}_{0.00} $ &$ 0.99^{+0.01}_{-0.16} $ &$ 0.98^{+0.02}_{-0.42} $ &$ 0.97^{+0.03}_{-0.64} $ &$ 0.95^{+0.05}_{-0.90} $ &$ 0.93^{+0.07}_{-0.89} $ &$ 0.91^{+0.09}_{-0.90} $  \\
Voyager & max & alltides&  $ 1.00^{+0.00}_{0.00} $ &$ 0.99^{+0.01}_{-0.13} $ &$ 0.98^{+0.02}_{-0.36} $ &$ 0.97^{+0.03}_{-0.56} $ &$ 0.95^{+0.05}_{-0.82} $ &$ 0.93^{+0.07}_{-0.86} $ &$ 0.92^{+0.08}_{-0.88} $  \\
Voyager & max & notides&  $ 1.00^{+0.00}_{0.00} $ &$ 0.99^{+0.01}_{-0.20} $ &$ 0.97^{+0.03}_{-0.51} $ &$ 0.96^{+0.04}_{-0.74} $ &$ 0.95^{+0.05}_{-1.18} $ &$ 0.93^{+0.07}_{-1.30} $ &$ 0.90^{+0.09}_{-1.23} $  \\
3$^{\rm rd}$ gen. & collapse & time&  $ 0.85^{+0.05}_{-0.64} $ &$ 0.82^{+0.07}_{-0.67} $ &$ 0.77^{+0.11}_{-0.70} $ &$ 0.72^{+0.15}_{-0.77} $ &$ 0.61^{+0.26}_{-0.78} $ &$ 0.55^{+0.31}_{-0.78} $ &$ 0.52^{+0.34}_{-0.72} $  \\
3$^{\rm rd}$ gen. & collapse & alltides&  $ 0.85^{+0.05}_{-0.63} $ &$ 0.82^{+0.07}_{-0.66} $ &$ 0.78^{+0.10}_{-0.69} $ &$ 0.73^{+0.14}_{-0.72} $ &$ 0.63^{+0.24}_{-0.68} $ &$ 0.56^{+0.30}_{-0.65} $ &$ 0.53^{+0.33}_{-0.67} $  \\
3$^{\rm rd}$ gen. & collapse & notides&  $ 0.85^{+0.05}_{-0.63} $ &$ 0.81^{+0.08}_{-0.70} $ &$ 0.74^{+0.14}_{-0.92} $ &$ 0.67^{+0.21}_{-1.19} $ &$ 0.53^{+0.33}_{-1.21} $ &$ 0.47^{+0.39}_{-1.12} $ &$ 0.47^{+0.38}_{-1.16} $  \\
3$^{\rm rd}$ gen. & uniform & time&  $ 0.50^{+0.34}_{-0.34} $ &$ 0.47^{+0.35}_{-0.36} $ &$ 0.43^{+0.37}_{-0.38} $ &$ 0.41^{+0.38}_{-0.46} $ &$ 0.36^{+0.40}_{-0.53} $ &$ 0.36^{+0.39}_{-0.55} $ &$ 0.34^{+0.41}_{-0.57} $  \\
3$^{\rm rd}$ gen. & uniform & alltides&  $ 0.50^{+0.34}_{-0.34} $ &$ 0.47^{+0.35}_{-0.36} $ &$ 0.44^{+0.36}_{-0.37} $ &$ 0.41^{+0.38}_{-0.40} $ &$ 0.38^{+0.39}_{-0.45} $ &$ 0.36^{+0.39}_{-0.48} $ &$ 0.35^{+0.39}_{-0.52} $  \\
3$^{\rm rd}$ gen. & uniform & notides&  $ 0.50^{+0.34}_{-0.35} $ &$ 0.46^{+0.36}_{-0.38} $ &$ 0.40^{+0.38}_{-0.51} $ &$ 0.37^{+0.40}_{-0.66} $ &$ 0.32^{+0.44}_{-0.74} $ &$ 0.30^{+0.45}_{-0.75} $ &$ 0.31^{+0.42}_{-0.72} $  \\
3$^{\rm rd}$ gen. & max & time&  $ 1.00^{+0.00}_{0.00} $ &$ 0.99^{+0.01}_{-0.36} $ &$ 0.97^{+0.03}_{-0.80} $ &$ 0.95^{+0.05}_{-1.01} $ &$ 0.91^{+0.09}_{-1.13} $ &$ 0.87^{+0.12}_{-1.20} $ &$ 0.86^{+0.14}_{-1.21} $  \\
3$^{\rm rd}$ gen. & max & alltides&  $ 1.00^{+0.00}_{0.00} $ &$ 0.99^{+0.01}_{-0.33} $ &$ 0.97^{+0.03}_{-0.74} $ &$ 0.95^{+0.05}_{-0.94} $ &$ 0.92^{+0.08}_{-0.97} $ &$ 0.89^{+0.11}_{-1.01} $ &$ 0.88^{+0.12}_{-1.04} $  \\
3$^{\rm rd}$ gen. & max & notides&  $ 1.00^{+0.00}_{0.00} $ &$ 0.98^{+0.02}_{-0.53} $ &$ 0.95^{+0.05}_{-1.20} $ &$ 0.92^{+0.08}_{-1.58} $ &$ 0.87^{+0.13}_{-1.73} $ &$ 0.85^{+0.15}_{-1.70} $ &$ 0.82^{+0.17}_{-1.70} $  \\
\end{tabular}
\caption{Medians of the marginalized effective-spin distributions in each of our model variations. Errors refer to the 5th and 95th confidence levels, respectively. All percentiles reported in this table have been weighted with the corresponding detection rates.}
\label{chieffpercentiles}
\end{table*}

We first illustrate our predictions for the marginalized distributions of $\chi_{\rm eff}$.
A summary of our findings is provided in Table~\ref{chieffpercentiles}, where we report detection-weighted medians and 90\% confidence intervals of $\chi_{\rm eff}$ for each of our simulations.

\begin{figure}
\includegraphics[width=\columnwidth]{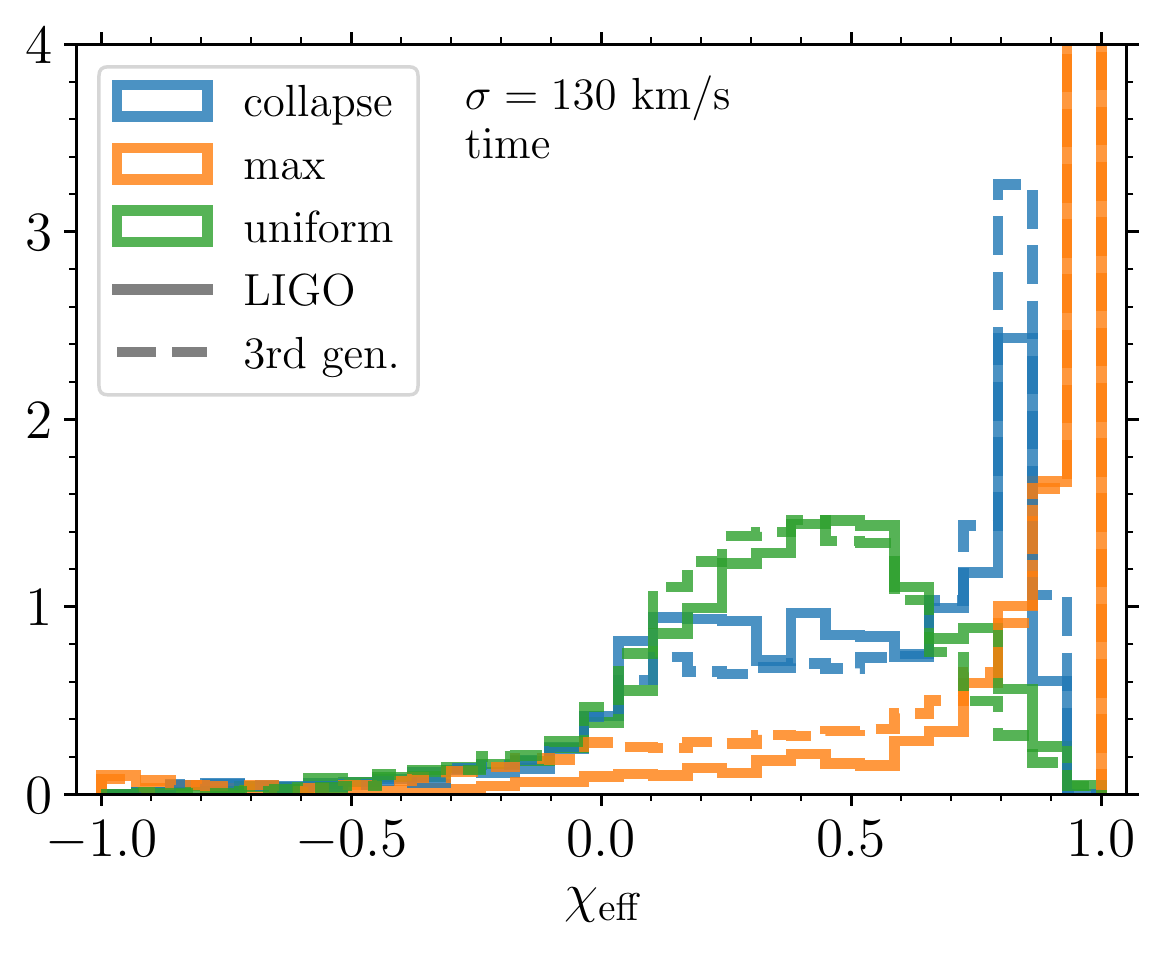}
\caption{Marginalized effective-spin distribution predicted by our three spin-magnitude models (\emph{collapse},\emph{max},\emph{uniform}) as detectable by LIGO and Cosmic Explorer. Results are shown for $\sigma=130$ km/s and the \emph{time} model of tidal interactions.}
\label{allspinschieff}
\end{figure}
Among our model variations, the spin magnitudes have the largest impact on  $\chi_{\rm eff}$. Figure~\ref{allspinschieff} shows the {detectable distributions of $\chi_{\rm eff}$ (cf. Sec.~\ref{detectability})} predicted by our three spin models, \emph{uniform}, \emph{collapse} and \emph{max}, assuming $\sigma=130$ km/s and the \emph{time} model of tidal interactions. The \emph{max} model predicts a sharp peak at $\chi_{\rm eff}=1$, while the  \emph{uniform} model presents a broader peak at $\chi_{\rm eff}\sim 0.5$. %
The \emph{collapse} case acts like a rough mixture of the two, with light BHs presenting preferentially large spins $\chi\!\sim\!0.9$, while the spins of heavier BHs can span a wider range. %

An interesting feature of these distributions is their dependence on the detector sensitivity. For the \emph{max} model, switching to a third-generation detector decreases the typical effective spin: the median in $\chi_{\rm eff}$ goes from $\sim\! 0.96$ to $\sim\!0.91$ (assuming the \emph{time}, $\sigma=130$ km/s distribution as shown in Fig.~\ref{allspinschieff}). The same is true for the
\emph{uniform} model, where the median decreases from $\sim\!0.41$ for LIGO to $\sim\!0.36$ for Cosmic Explorer. 
The orbital hangup effect (cf. Sec.~\ref{detectability}) causes a selection bias on GW measurements: binaries with negative (positive) $\chi_{\rm eff}$ have a shorter (longer) waveform and therefore are harder (easier) to detect  \cite{2018arXiv180503046N}. A detector with better sensitivity reduces this selection bias, thus pushing the median of the detectable events to lower values. %

The \emph{collapse} model behaves in the opposite way: better instruments will detect larger $\chi_{\rm eff}$'s (median increasing from $\sim\!0.49$ to $\sim\!0.61$). This is because instrumental improvements in the high-frequency range will make us sensitive to lower mass systems which, in the \emph{collapse} model, have preferentially high spins (cf. Fig.~\ref{fitspin}). The hangup effect is still present, but turns out to be subdominant.

\begin{figure}
\includegraphics[width=\columnwidth]{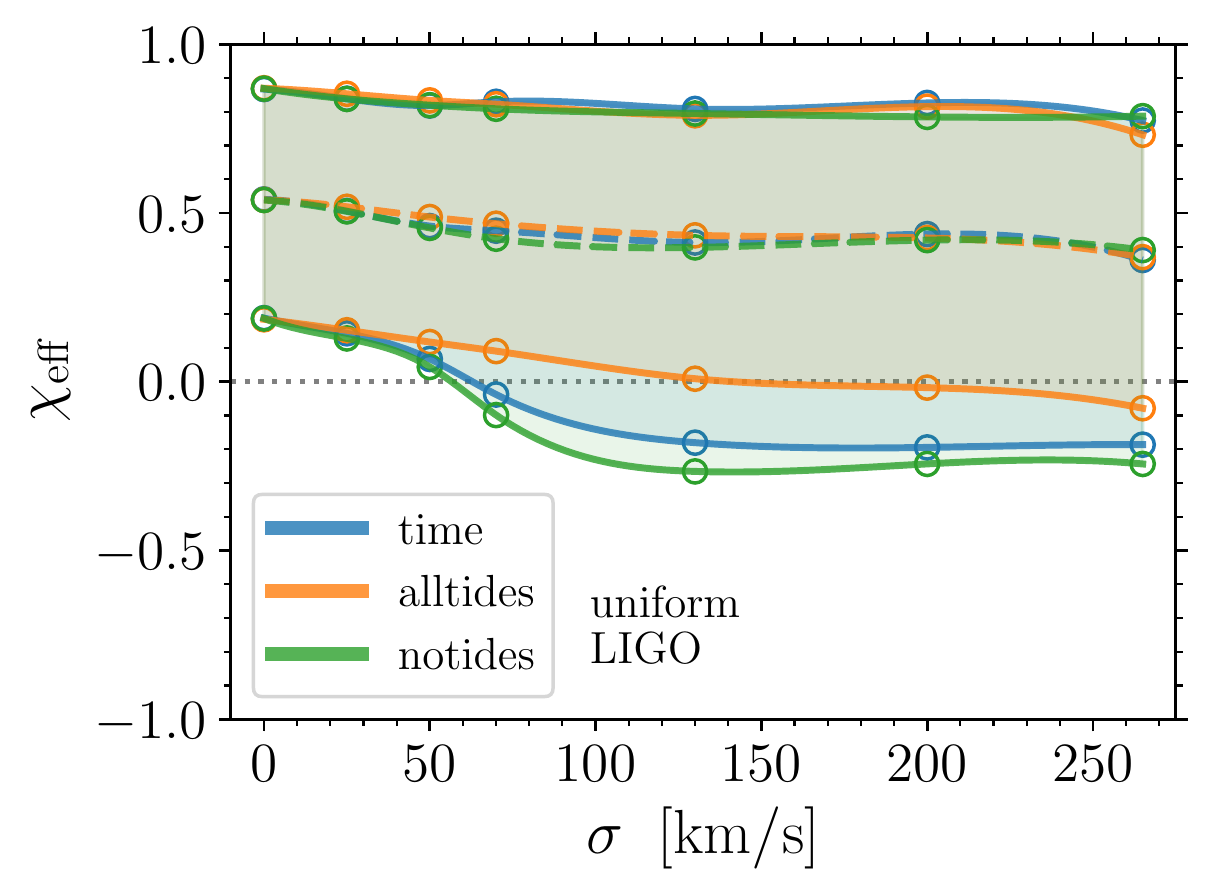}
\caption{Median values (dashed lines) and 90\% confidence intervals (solid lines and shaded areas) for $\chi_{\rm eff}$ as a function of natal kicks and tidal interactions (\emph{time}, \emph{alltides}, \emph{notides}). Results are shown for \emph{uniform} spin magnitudes and weighted with LIGO detection rates.}
\label{alltideschieff}
\end{figure}
Figure~\ref{alltideschieff} illustrates the effect of tidal interactions on the predicted values of $\chi_{\rm eff}$. In our models tides only affect the spin orientations and, as expected, produce larger $\chi_{\rm eff}$ values.
Within the context of these models, tides are less important than natal spins to predict the effective-spin distributions. Notably, tides mainly affect the
 small-$\chi_{\rm eff}$
 tail of the population. As illustrated at length below, negative values of $\chi_{\rm eff}$ are hard to explain in the \emph{alltides} model (where all stellar spins are realigned in between the two explosions), while they are relatively easy to accommodate with both the \emph{notides} and the \emph{time} models.

\subsection{On the sign and symmetry of $\chi_{\rm eff}$}

From Eq.~(\ref{chieff}), it is obvious that only largely misaligned spins can produce negative values of $\chi_{\rm
  eff}$.  It has been suggested that a single
 confirmed measurement of a system with $\chi_{\rm eff}<0$ could rule out isolated BH formation
for that event
  in favor of dynamical interactions \cite{2016ApJ...832L...2R}. One of the events observed so far (GW151226
  \cite{2016PhRvL.116x1103A}) has $\chi_{\rm eff}>0$ at very high confidence. Some of the other events present more
  posterior weight at negative $\chi_{\rm eff}$ values, but $\chi_{\rm eff}\geq0$ cannot be ruled out. We stress that
  these significance assessments  have to be taken with care as 
 they depend on the Bayesian prior used in the analysis \cite{2017PhRvL.119y1103V}.

\begin{figure}
\includegraphics[width=\columnwidth]{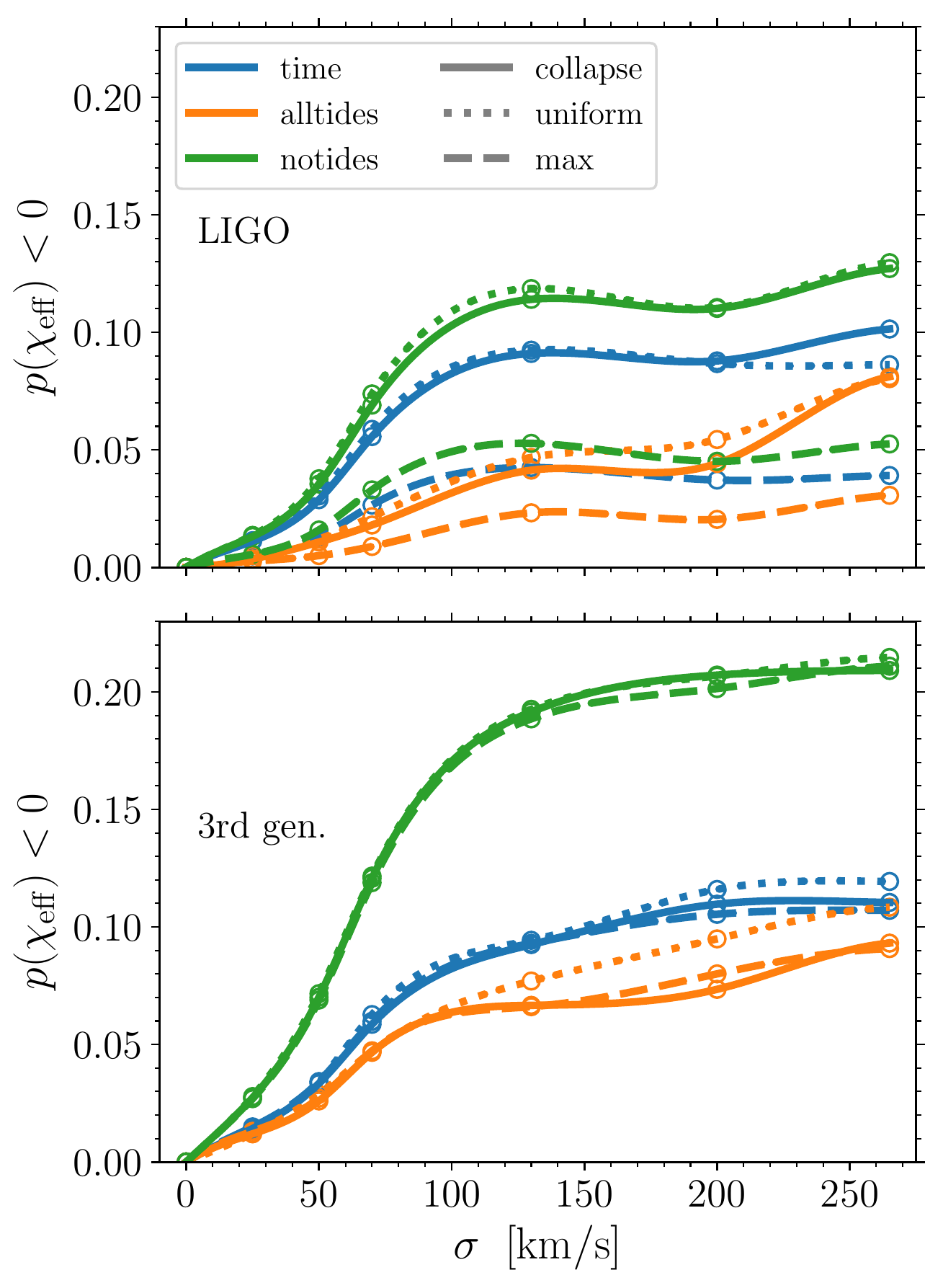}
\caption{Fraction of binaries  with negative effective spin as a function of natal kicks ({\it x} axis), tidal interaction (colors) and spin-magnitude model (line styles). Top (bottom) panels shows results for LIGO (Cosmic Explorer).}
\label{negativechieff}
\end{figure}

Figure~\ref{negativechieff} shows the predicted rate fraction of BH binaries with $\chi_{\rm eff}<0$ in each of our models. As expected, misalignments are larger for larger kicks and, consequently, $p(\chi_{\rm eff}<0)$ increases as $\sigma$ increases.  The typical fraction of binaries with negative effective spins detectable by LIGO ranges from $\sim3\%$ to $\sim10\%$ (with the exception of $\sigma=0$ km/s, where $\chi_{\rm eff}\geq 0$ by construction).

Our results show that isolated pairs of stars  \emph{can} explain single events with $\chi_{\rm eff}<0$, in disagreement with the main claim made by \cite{2016ApJ...832L...2R} (but see their Fig. 3). Obviously, since we are assuming that stars are initially aligned with the orbital angular momentum, BH spins cannot be misaligned if kicks are not present. The fiducial model of \cite{2016ApJ...832L...2R} heavily suppresses kicks for BHs compared to neutron stars, thus effectively preventing misalignments. A more conservative statement is the following: single detections with $\chi_{\rm eff}<0$ would point towards dynamical interaction, \emph{if} stellar spins are initially aligned and BH kicks are heavily suppressed. Even moderate kicks of $\sigma=25$ km/s allow for $p(\chi_{\rm eff}<0)\sim 3\%$.

Together with kicks, tidal interactions are important to determine the sign of $\chi_{\rm eff}$. Higher (lower) fractions of negative effective spins are predicted for the \emph{notides} (\emph{alltides}) model, while the \emph{time} model lies somewhere in between. 
Notably, $p(\chi_{\rm eff}<0)$ is largely independent of the spin-magnitude assumption (especially for third-generation detectors). 

This suggests that, at least in the context of well-specified astrophysical models like ours, $\chi_{\rm eff}$ measurements alone can partially break the degeneracy between spin magnitudes and spin  directions encoded in Eq.~(\ref{chieff}). Natal spins mainly determine the broad shape of the distribution (Fig.~\ref{allspinschieff}), while alignment processes have a clean impact on the low-$\chi_{\rm eff}$ tail (Fig.~\ref{negativechieff}).

Although negative values of $\chi_{\rm eff}$ are possible, our distributions are far from being symmetric (cf. e.g. Table~\ref{chieffpercentiles} where all medians are $\gtrsim 0.3$). On the contrary, dynamical formation channels predict spins isotropically distributed (although see \cite{2017NatAs...1E..64C}), which corresponds to a marginalized effective-spin distribution symmetric about $\chi_{\rm eff}=0$. Our models suggest that the symmetry of the $\chi_{\rm eff}$ is a robust indicator to distinguish isolated binary formation from dynamical interactions. We therefore confirm the ideas put forward  by \cite{2017Natur.548..426F,2018ApJ...854L...9F} with large-scale population-synthesis simulations.

\begin{figure*}[t]
\includegraphics[width=\textwidth]{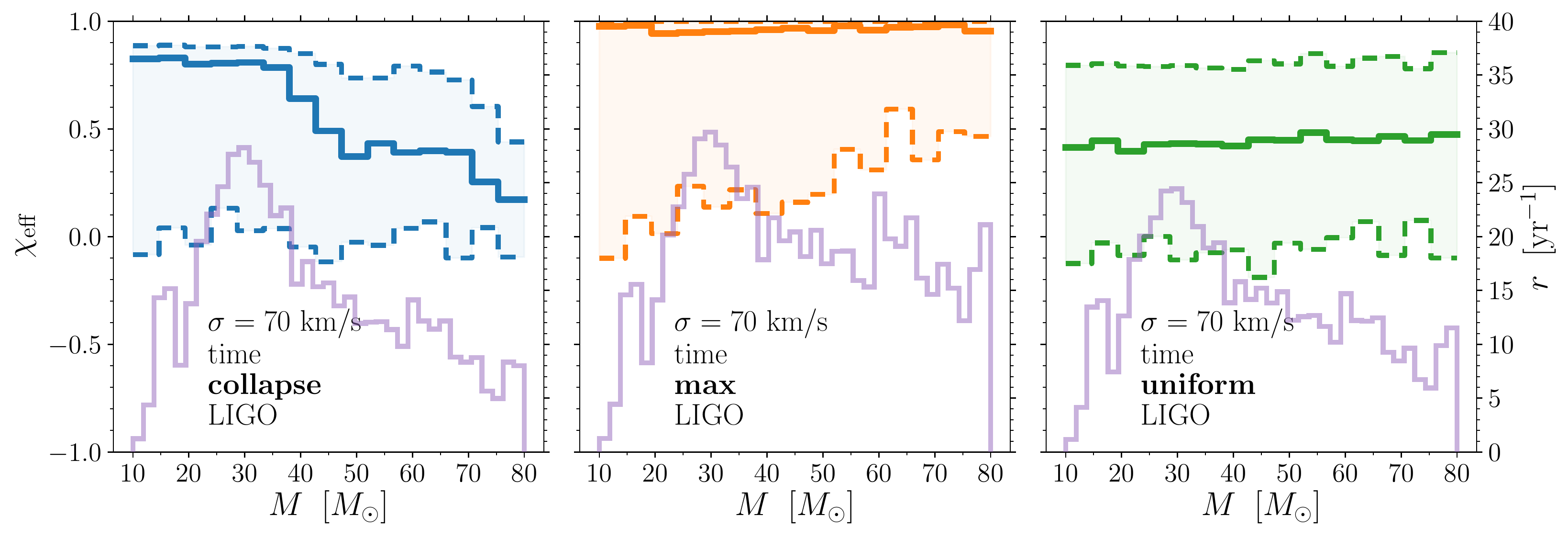}
\caption{Effective spins $\chi_{\rm eff}$ and detection rates $r$  in bins of total mass $M$. We show results for our three spin models (left: \emph{collapse}; middle: \emph{max}; right: \emph{uniform}) assuming $\sigma=70$ km/s, the \emph{time} model for tides, and the LIGO sensitivity curve. Thick solid (dashed) lines show median (90\% confidence interval) of $\chi_{\rm eff}$, as reported on the right {\it y} axis. Light histograms show the cumulative detection rates  in each mass bin, as  reported on the right y-axis.}
\label{masschieff}
\end{figure*}
\begin{figure}[h!]
\includegraphics[width=\columnwidth]{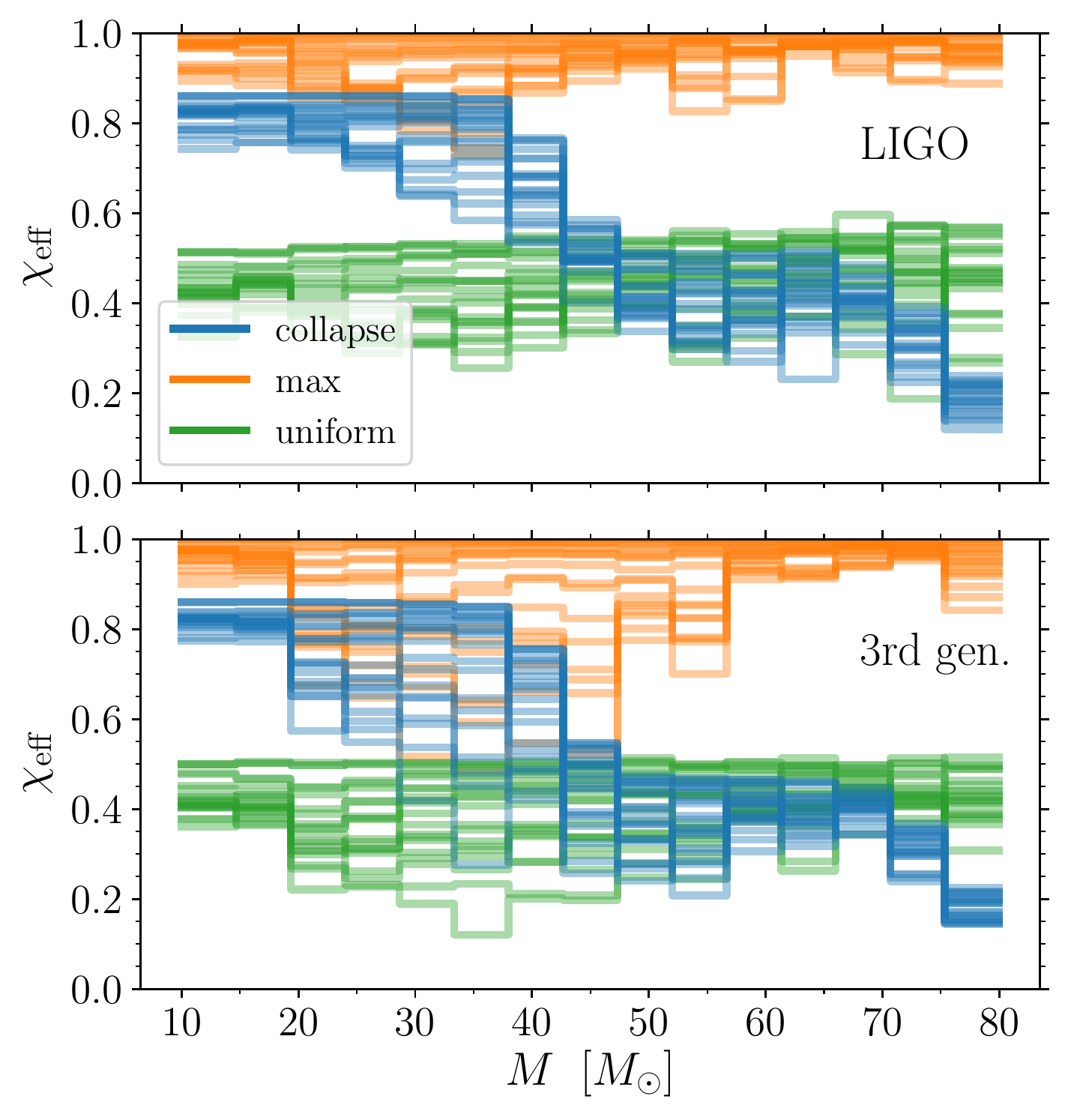}
\caption{Medians of $\chi_{\rm eff}$ as a function of the total mass $M$ for all of our model variations. The top panel shows results weighted by LIGO detection rates, while the bottom panel assumes a third-generation detector (Cosmic Explorer). Colors differentiate our three spin-magnitude models (\emph{collapse}: blue; \emph{max}: orange; \emph{uniform}: green). In each series, the various lines are obtained by varying over tidal interactions (\emph{time}, \emph{alltides}, \emph{notides})  and kick magnitudes ($\sigma=0,25,50,70,130,200, 265$ km/s).}
\label{masskicks}
\vspace{1cm}
\end{figure}

\subsection{Mass dependence}

In Fig.~\ref{masschieff} we present predictions  for the effective spins of BH binaries with different total source-frame masses. Results for the  \emph{collapse} model directly reflect the injected relationship between BH masses and spins.  In the absence of this correlation, a simpler trend emerges, namely that kicks more easily misalign light systems. This is especially evident in the \emph{max} case because all BH spin magnitudes are equal.

Figure~\ref{masschieff} also shows the expected detection rates as a function of $M$. The three panels are constructed with the very same \textsc{StarTrack} evolution ($\sigma=70$ km/s), which predicts a  mass spectrum peaking at about $M\!\sim\! 30 M_\odot$. This is the strongest feature visible in all three distributions shown in Fig.~\ref{masschieff}.

Differences in $r$ between the three panels are a direct consequence of the spinning waveform model used to compute the horizon distance. This effect was mostly neglected in previous \textsc{StarTrack} studies, with the exception of \cite{2015ApJ...806..263D}. 
For partially aligned systems like ours, the orbital hangup effect facilitates the detection of BH binaries with large spin magnitudes. The \emph{max} rates are therefore higher than those predicted by the \emph{uniform} model. In model \emph{collapse}, where heavy BHs spin slower compared to less massive ones,  detection rates at large  (low) $M$ are suppressed (enhanced). As expected, the behavior changes around $M\sim 40 M_\odot$, which is roughly twice the value of the turnover of Fig.~\ref{fitspin}.

Fig.~\ref{masskicks} shows medians of $\chi_{\rm eff}$ as a function of $M$ for all kick, spin and tide variations. The main takeaway here is that distributions are qualitatively very similar for all models of tidal interactions and natal kicks, and only depend on the spin-magnitude variation. This finding further stresses one of the points made above: $\chi_{\rm eff}$ measurements alone can provide powerful constraints on BH natal spins, even in the presence of misalignment processes.

\begin{figure*}
\includegraphics[width=\textwidth]{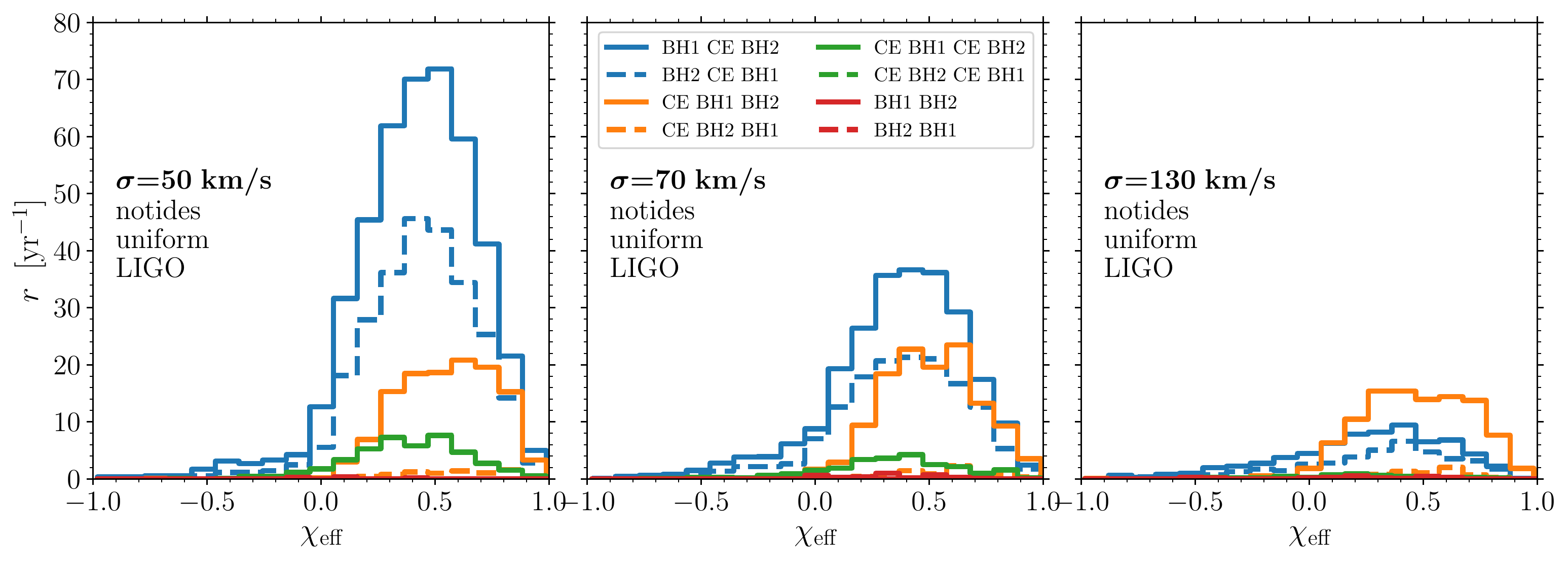}
\caption{Detection rates in bins of effective spin $\chi_{\rm eff}$ divided into formation channels. Here ``BH1'' and ``BH2'' stand for the formation of the heavier and lighter BH, respectively, while ``CE'' stand for the occurrence of a common-envelope phase (cf. Sec.~\ref{pathway}). Results are shown for the \emph{notides}, \emph{uniform} spin model as detectable by LIGO. Other models have qualitatively similar results. Natal kicks are varied in the three panels as indicated in the legend.}
\label{splitformchannels}
\end{figure*}

\subsection{Constraints on formation channels}

Finally, we present our predictions for $\chi_{\rm eff}$ in the eight different formation channels introduced in Sec.~\ref{pathway}. Fig.~\ref{splitformchannels} shows results for some of our \emph{uniform}, \emph{notides} models. It is illustrative to look at this variation in particular because, as described above (cf. Figs.~\ref{allspinschieff} and \ref{alltideschieff}), it maximizes the fraction of binaries with $\chi_{\rm eff}$ far from unity. We stress, however, that the trends described here are illustrative of all of our distributions.

As already shown in Fig.~\ref{ratechannels}, the fraction of ``standard'' binaries with a common-envelope evolution in between the two supernovae decreases with the kick-velocity dispersion parameter $\sigma$.  For $\sigma\gtrsim 100$ km/s kicks unbind most binaries at the first supernova, unless the binary separation was already tight because of an earlier common-envelope phase. Binaries in those channels (see in particular ``CE BH1 BH2'' in Fig.~\ref{ratechannels}) are largely unaffected by  kicks. Their orbital angular velocity is so large that they not only remain bound, but also  roughly aligned.

Largely misaligned binaries all belong to the more standard ``BH1 CE BH2'' and ``BH2 CE BH1'' channels, independently of $\sigma$. %
This result illustrates a clean prediction of our models: binaries with small $\chi_{\rm eff}$ are formed following very specific pathways, namely those which present a common-envelope phase \emph{between} the formation of the two BHs (and not earlier).

This observation can be rephrased as follow: if kicks are large, binaries in the  ``BH1 CE BH2'' and  ``BH2 CE BH1'' channels are either unbound or, if they survive, they are largely misaligned. This behavior can be explained with some simple kinematics.
For a circular orbit, the spin misalignment angle $\theta$ imparted by a kick is \citep{2000ApJ...541..319K,2017PhRvL.119a1101O}
\begin{equation}
\cos\theta = \frac{ |\mathbf{v}|+\mathbf{v_k} \cdot \mathbf{\hat v}}{ \sqrt { (|\mathbf{v}|+\mathbf{v_k} \cdot \mathbf{\hat v})^2 + (\mathbf{v_k}\cdot \mathbf{\hat L})^2 }},
\end{equation}
where $\mathbf{v_k}$ is the kick velocity, $\mathbf{v}$ is the orbital velocity and $\mathbf{\hat L}$ is the direction of the orbital angular momentum before the explosion.
Since $\mathbf{v_k}$ is drawn from a Maxwellian distribution, the component $\mathbf{v_k} \cdot \mathbf{\hat v}$  and $\mathbf{v_k} \cdot \mathbf{\hat L}$ are Gaussian. In the limit of large $\sigma$, this implies that $\theta$ 
 is uniformly distributed \citep{marsaglia1972}. As the kick increases, more binaries are unbound while the distribution of misalignments flattens.

\section{Results: spin directions}
\label{results2}

We now explore predictions of our models for the individual directions of the two spins. %
As already outlined in Sec.~\ref{methods}, the mutual orientations of the two spins and the orbital angular momentum can be described by three variables:
$\theta_1$ and $\theta_2$ are the angles between the two spins and the orbital angular momentum, and $\Delta\Phi=\phi_2-\phi_1$ is the angle between the projections of the two spins onto the orbital plane (see e.g. Fig.~1 of \cite{2015PhRvD..92f4016G} for a schematic representation). %
The angles $\theta_1$ and $\theta_2$ are polar angles, defined in the range $[0,\pi]$, while $\Delta\Phi$ is an azimuthal angle defined in the range $[-\pi,\pi]$. However, precession cycles are symmetric in PN dynamics \cite{2015PhRvD..92f4016G}, so that we can consider $\Delta\Phi\in [0,\pi]$ without loss of generality. 

The punch line of this section (which generalizes the toy model of \cite{2013PhRvD..87j4028G} to state-of-the-art astrophysical populations) is that BH spin orientations near merger fall into three well-separated subpopulations. These classes of BH binaries carry the imprint of specific physical processes driving the evolution of their stellar progenitors.

\subsection{Evolution of the spin tilts}

\begin{figure*}[p]
\centering
\includegraphics[width=0.46\textwidth,page=1]{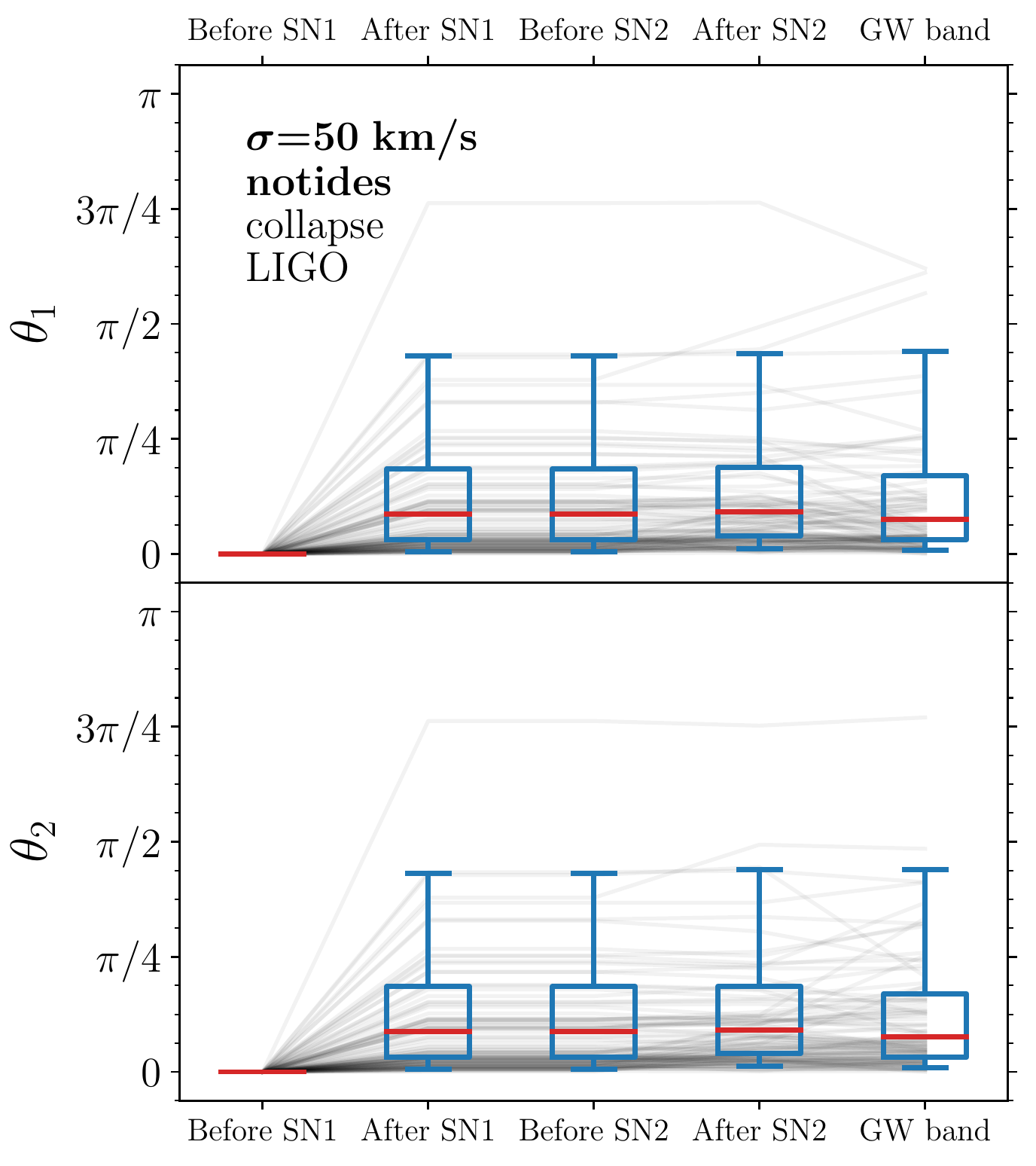}
\hspace{0.03\textwidth}
\includegraphics[width=0.46\textwidth,page=2]{violin}\\
\vspace{0.6cm}
\includegraphics[width=0.46\textwidth,page=3]{violin}
\hspace{0.03\textwidth}
\includegraphics[width=0.46\textwidth,page=4]{violin}
\caption{Evolution of the spin orientations along the lives of BH-binary progenitors detectable by LIGO. The top (bottom) subpanel in each plot shows the tilt $\theta_1$ ($\theta_2$) of the object forming the more (less) massive BH. All binaries are aligned before the first supernova (SN1), which imparts a first tilt to both spins. Tidal interactions can realign one of the spins in between the two explosions. The second kick (SN2) sets the spin misalignment angles at BH-binary formation. These orientations then evolve under the influence of relativistic spin-spin and spin-orbit couplings until they become detectable in GWs (roughly at $f_{\rm GW}=20$ Hz). At each stage, the median of the distribution is marked with a red line; the blue boxes (bars) include 50\% (90\%) of the detection rate. Thin gray lines show individual evolutionary tracks for the 100 binaries with the highest detection rates in each sample.}
\label{violin}
\end{figure*}

First, we illustrate the evolution of the spin angles during the various steps of binary stellar evolution. There are five key stages where spin directions can change. These are listed below and illustrated in Fig.~\ref{violin}, where we track changes of the two tilt angles along each stage by separating progenitors that form the more ($\theta_1$) and less ($\theta_2$) massive BHs.

\begin{enumerate}
\item
Our initial assumption is that primordial misalignments are negligible, i.e. $\theta_1=\theta_2=0$ at the beginning of each evolution.

\item
The first stage where spin tilts can change is the supernova that forms the first BH (SN1). This is typically, but not always, the collapse event where the more massive BH is formed (c.f  Sec.~\ref{pathway}). It turns out that the kick imparted at the first explosion is the dominant effect setting the spin directions in the entire evolution, and all other stages play a subdominant role (cf. \cite{2017PhRvL.119a1101O}, where this consideration was used to estimate $\sigma$ from GW151226 data). On average, larger kicks introduce larger misalignments. However, this trend is mitigated by the fact that larger kicks also unbind binaries. Only the harder binaries in the sample survive strong kicks, and those same binaries are harder to tilt. At this stage, the median in the angles $\theta_1=\theta_2$ (both members receive the same tilt) is $\sim \pi/8$, and this number changes only weakly among our kick and spin models. The large-misalignment tail of the tilt distributions, on the other hand, depends strongly on $\sigma$: tilts $\theta_i\gtrsim \pi/2$ require $\sigma\gtrsim 50$ km/s. This is consistent with the results already presented in Fig.~\ref{negativechieff}, where indeed curves steepen at about $\sigma\!\sim\!50$ km/s.

\item
After the first explosion, the system is formed by a BH and a star. At this stage, tidal interactions can realign the stellar spin. All stars are realigned in the  \emph{alltides} case where, consequently, one of the two spin misalignment angles drops to zero. In the majority of the cases, tides enforce $\theta_2=0$ between the two explosions, because the first explosion typically forms the most massive BH. However, if the less massive BH is formed first, tidal alignment enforces $\theta_1=0$. Both spin misalignments are unchanged in the \emph{notides} models, where tidal realignment is assumed to be completely inefficient.

\item 
  The second supernova (SN2) imparts another tilt to the orbital plane. Before this second kick, the binary already underwent a common-envelope phase which greatly tightened the separation.\footnote{Cases ``BH1 BH2'' and ``BH2 BH1'' of Sec.~\ref{pathway} are an exception but their rates  are extremely low in all our models}
  Because of their larger orbital velocities, binaries are much harder to tilt at this stage compared to the first explosion. This second kick is virtually irrelevant for $\sigma\lesssim 100$ km/s. In the case of larger kicks where the ``CE BH1 BH2'' channel dominates, on average the second explosion  increases the misalignment angle (this is trivially true for the spin that was previously realigned by tides). After the second kick, the tilt angles $\theta_1$ and $\theta_2$ are in general not equal to each other. Even in the $\emph{notides}$ cases where $\theta_1=\theta_2$ before the SN, spin precession might affect the azimuthal angles ($\phi_1\neq \phi_2$ if $t_{\rm SN}>t_{\rm pre}$, cf. Sec.~\ref{methods}) and consequently the post-SN tilts.
  
\item Finally,  PN evolutions   \cite{2016PhRvD..93l4066G} are used to propagate binaries from BH formation (after SN2) to detection, here assumed to happen when the GW emission frequency drops below $f_{\rm GW}=20$ Hz. PN evolutions could last Gyrs, where binaries undergo many precession cycles. The tilt angles are modified by relativistic spin-spin and spin-orbit couplings. These are conditioned to keep $\chi_{\rm eff}$ constant \cite{2008PhRvD..78d4021R,2015PhRvD..92f4016G}, such that $\cos\theta_1$ increases only when $\cos\theta_2$ decreases, and vice versa.

\begin{figure*}
\includegraphics[width=\textwidth]{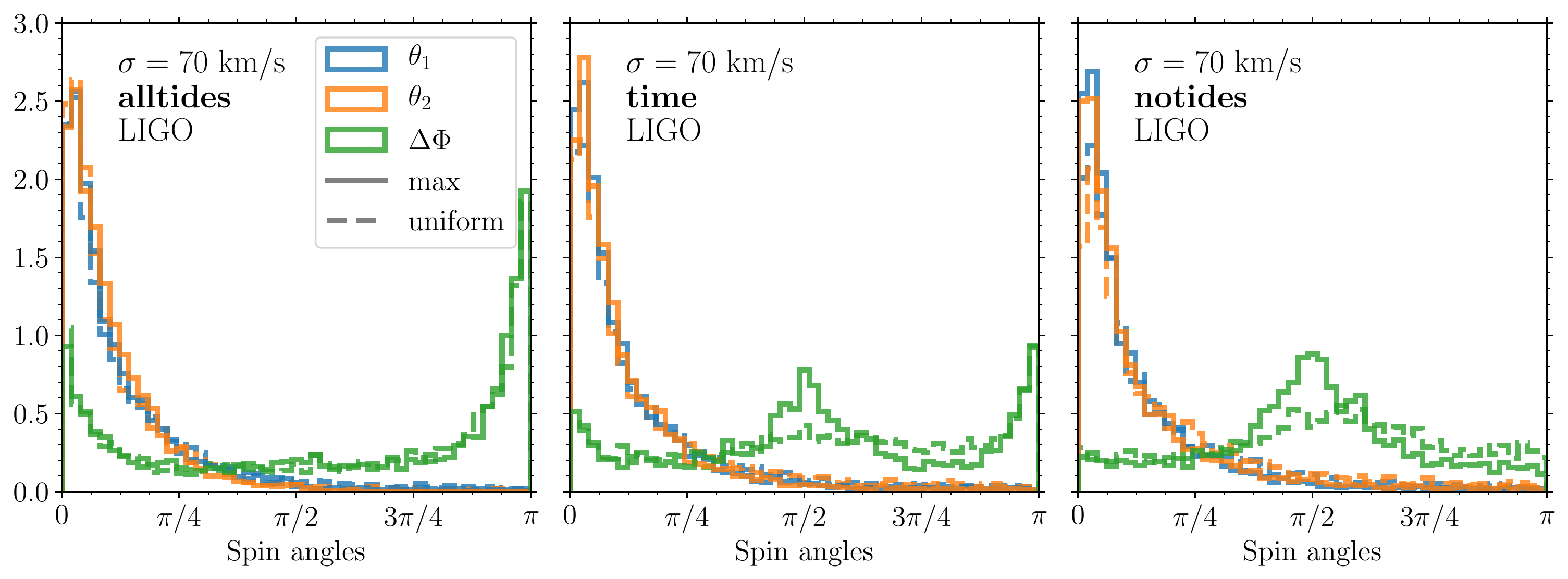}
\caption{Marginalized distributions of the spin angles $\theta_1$, $\theta_2$ and $\Delta\Phi$ at $f_{\rm GW}=20$ Hz for binaries detectable by LIGO. We assume natal kicks of $\sigma=70$ km/s and a variety of assumptions on tides and spin magnitudes at collapse. Distributions of $\theta_1$ and $\theta_2$ are  peaked at 0, with widths $\sim \pi/4$. Distributions of $\Delta\Phi$ at detection carry clear imprints  of the underlying assumptions on tidal interactions and realignment. }
\label{angles20hz}
\end{figure*}

\end{enumerate}

We now explore in detail the spin orientations at the last stage, when binaries become detectable in GWs.

\subsection{Spin angles at detection}

Figure \ref{angles20hz} shows marginalized distributions of $\theta_1$, $\theta_2$ and $\Delta\Phi$ at $f_{\rm GW}=20$ Hz for a subset of our models. The tilt angle distributions peak at $\theta_1\simeq 0 $ and $\theta_2\simeq 0$, which is the initial assumption for all stars in our models. Misalignments are introduced by natal kicks. The typical widths of the $\theta_1$ and $\theta_2$ distributions vary from $\sim \pi/8$ for $\sigma=25$ km/s to  $\sim \pi/4$ for $\sigma=265$ km/s. These curves are largely independent of the chosen spin model.

On the contrary, the behavior of the angle $\Delta\Phi$ strongly depends on the spin variation. %
If all binaries are subject to tidal realignment (\emph{alltides}), the distribution of $\Delta\Phi$ at 20 Hz is strongly peaked at $\Delta\Phi=0,\pi$, while a less prominent peak at $\Delta\Phi=\pi/2$ is present if tides are suppressed (\emph{notides}). If only some of the binaries are realigned (\emph{time}), three distinct  subpopulations are present, which pile up at $\Delta\Phi=0,\pi/2$ and $\pi$.
Assumptions on the spin magnitude also play a visible role in Fig.~\ref{angles20hz}. PN spin-spin and spin-orbit couplings are weaker for lower spins and the peaks in $\Delta\Phi$ are consequently less pronounced. It is worth noting, however, that the impact of tides can be clearly seen even for the \emph{uniform} spin-magnitude model. %

The reason for this peculiar behavior of the precessional phase $\Delta\Phi$ lies in the PN evolution that binaries undergo between formation and detection. Spin precession naturally separates populations that formed with different tilt angles $\theta_1$, $\theta_2$ into different distributions for $\Delta\Phi$ . This is a well- known PN effect, first discovered by Schnittman \cite{2004PhRvD..70l4020S} and later explored in detail by \cite{2010PhRvD..81h4054K,2010ApJ...715.1006K,2012PhRvD..85l4049B,2013PhRvD..87j4028G,2015PhRvD..92f4016G} (see also \cite{2018arXiv180307695A,2017PhRvD..96b4007Z,2017CQGra..34f4004G,2016MNRAS.457L..49C,2014CQGra..31j5017G} for later investigations).  
Tidal interactions affect whether the tilt angle of the second formed BH is set to zero between the two explosions, thus strongly impacting its tilt angle at formation. As the binary inspirals towards merger, PN spin evolution tends to mix up the $\theta_1$ and $\theta_2$ distributions and separate the subpopulations in the variable $\Delta\Phi$. %

\begin{figure*}
\includegraphics[width=\textwidth]{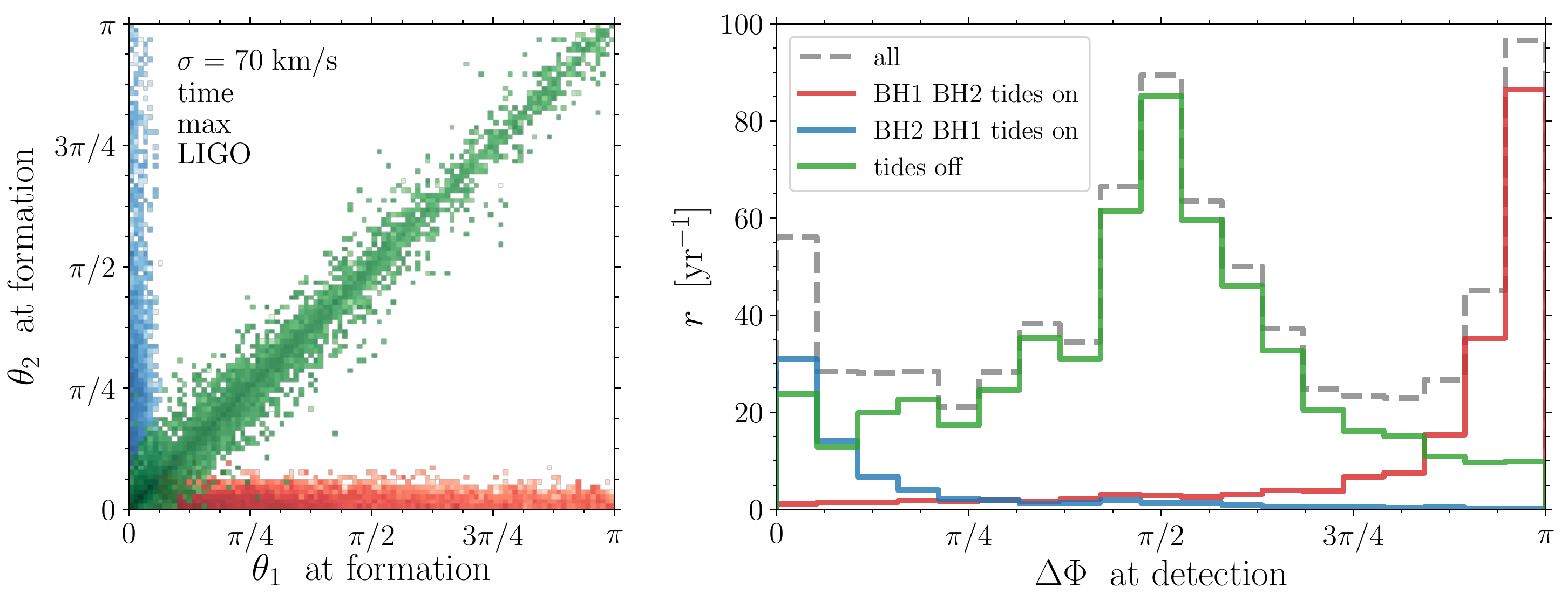}
\caption{Spin angles for three subchannels describing whether tidal realignment was efficient or not (``tides on'' vs. ``tides off'') and whichever BH formed first (``BH1 BH2'' vs ``BH2 BH1''). The left panel shows a two-dimensional histogram of the tilt angles $\theta_1$ and $\theta_2$ at BH formation, where the three subpopulations are clearly separated. Those same binaries are evolved to detection ($f_{\rm GW}=20$ Hz) and the variable $\Delta\Phi$ at that point is shown in the right panel. PN evolution naturally clusters binaries from different sub-channels to ``orthogonal'' regions in the $\Delta\Phi$ parameter space.
This figure was generated assuming the LIGO sensitivity, $\sigma=70$ km/s, the \emph{time} tidal alignment model and the \emph{max} spin model.}
\label{deltaphisplit}
\end{figure*}

To better illustrate and quantify this behavior, let us divide our binaries in three subpopulations. First, we select binaries that were not realigned by tidal interactions between the two explosions (``tides off''). This will correspond to $100\%$ (0\%) of the binaries in the \emph{notides} (\emph{alltides}) models, and to some other fractions in the \emph{time} cases. For the remaining binaries that do undergo tidal realignment (``tides on''), we track whether BH1 forms before/after BH2 (thus grouping the eight channels of Sec.~\ref{pathway} into two). This results in three mutually exclusive subchannels: %
\begin{center}
\begin{tabular}{l@{\hskip 0.5in}l}
{\bf 1.}$\;\;\;$BH1 BH2 tides on \\
{\bf 2.}$\;\;\;$BH2 BH1 tides on\\
{\bf 3.}$\;\;\;$tides off
\end{tabular}
\end{center}

Figure~\ref{deltaphisplit} shows the resulting distribution of the spin angles. Crucially, we pair distributions of $\theta_1$ and $\theta_2$ at BH formation to the distribution of $\Delta\Phi$ at detection. Tides and the order of BH formation strongly separate the tilt angle distributions: $\theta_1\simeq\theta_2$ if tidal realignment is prevented, while $\theta_1\simeq 0$ ($\theta_2\simeq0 $) if tides are present and the realigned star ends up forming the primary (secondary) BH. Because of the long PN evolution before merger \cite{2015PhRvL.114h1103K,2015PhRvD..92f4016G,2016PhRvD..93l4066G}, these three populations are found 
with preferential values of $\Delta\Phi$ 
as they enter the LIGO band: 
$\Delta\Phi\sim 0$ for ``BH2 BH1 tides on'', $\Delta\Phi\sim \pi/2$ for ``tides off'' and $\Delta\Phi\sim\pi$ for ``BH1 BH2 tides on''.

These findings  confirm the toy model developed by some of the present authors \cite{2013PhRvD..87j4028G}, which only considered a few fiducial sources (compare e.g. their
Fig.~1 to Fig.~\ref{deltaphisplit} in this paper).    Despite a substantial extension to a
much larger population, including state-of-the-art initial conditions provided by \textsc{StarTrack}, and more complex models for
tidal alignment, this simple approach provides an accurate description of several key features of the population of spinning
binaries. 

Motivated by the identification of these three, coarsely identified classes, in the next section we employ
another tool developed in  \cite{2015PhRvL.114h1103K,2015PhRvD..92f4016G} to characterize precessing sources: their  \emph{spin morphology}.

\section{Results: spin morphologies}
\label{results3}

The spin morphology is a better tool to quantify spin precession in merging BH binaries. It was first introduced by \cite{2015PhRvL.114h1103K,2015PhRvD..92f4016G} (see also \cite{2018JPhCS.957a2014G} for a concise introduction). Here we briefly review the main concepts behind spin morphology, and then explore the implications for our populations of detectable BH binaries.

\subsection{A slowly evolving feature}

\begin{figure}
\centering
\includegraphics[width=\columnwidth]{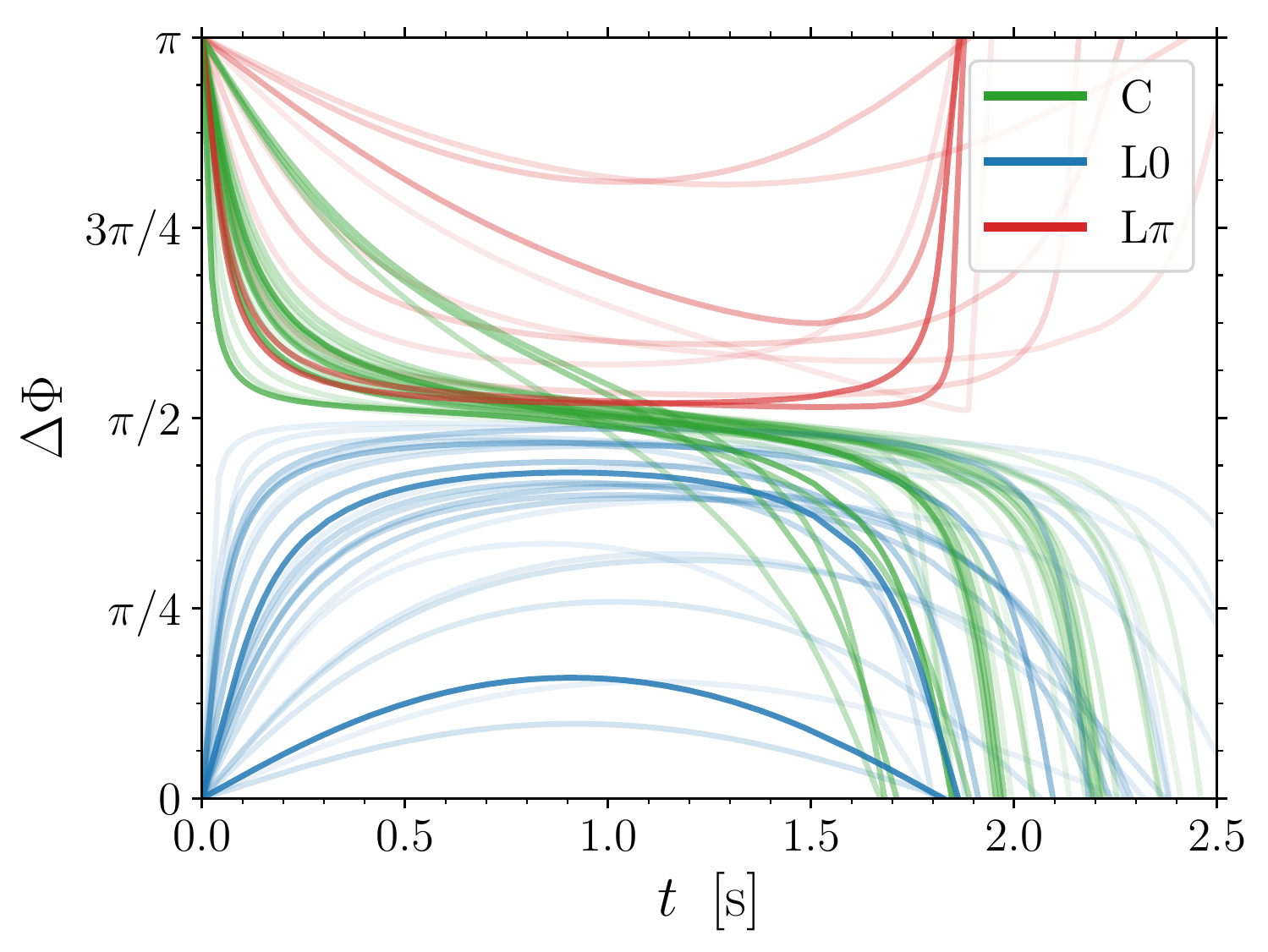}
\caption{Evolution of the angle $\Delta\Phi$ during (half of) a precession cycle at $f_{\rm GW}=20$ Hz for a sample of detectable BH binaries. 
We select some binaries among those with higher detection rates from our model with $\sigma=70$ km/s, the \emph{time} tidal alignment model and the \emph{max} spin magnitude model (as in Fig.~\ref{deltaphisplit}). Each line is shaded according to the LIGO detection rate of the corresponding source.
}
\label{explainmorphology}
\end{figure}

The qualitative shape of the precession cones of the two spins and the orbital angular momentum can be classified into three mutually exclusive classes based on the evolution of the precessional phase $\Delta\Phi$. 
Figure~\ref{explainmorphology} shows the evolution of $\Delta\Phi$ during single precession cycles at $f_{\rm GW}=20$ Hz for some indicative binaries from our distributions. 
For simplicity, we define a precession cycle to start and end at configurations where the three vectors $\mathbf{S_1}$, $\mathbf{S_2}$ and $\mathbf{L}$ are coplanar. These correspond to either $\Delta\Phi=0$ \emph{or} $\Delta\Phi=\pi$. There are  therefore three discrete possibilities.
\begin{enumerate}
\item Both configurations $\Delta\Phi=0$ and $\Delta\Phi=\pi$ are allowed, and the angle $\Delta\Phi$ circulates  in the full range $[0,\pi]$ during each precession cycle (C).
\item The configuration $\Delta\Phi=\pi$ is forbidden; the precession cycle consists of librations about $\Delta\Phi=0$ (L$0$).
\item The configuration $\Delta\Phi=0$ is forbidden; the precession cycle consists of librations about $\Delta\Phi=\pi$  (L$\pi$).
\end{enumerate}

This classification elucidates the results already presented in Figs.~\ref{angles20hz} and \ref{deltaphisplit}. Binaries in the two librating morphologies L$0$ and L$\pi$ spend more time close to the coplanar configurations, and are thus more likely to be found with either $\Delta\Phi=0$ or $\Delta\Phi=\pi$, respectively. Binaries in the circulating morphology C behave in the opposite way. For these binaries, the ``azimuthal velocity'' $d\Delta\Phi /dt$ is larger at $\Delta\Phi\simeq0,\pi$ and  lower at $\Delta\Phi\simeq \pi/2$. Sources naturally spend more time where $d\Delta\Phi /dt$ is lower, and are thus more likely to be found with $\Delta\Phi\simeq \pi/2$.

The most notable feature about the spin morphology is its slow variation. In BH-binary systems, spins vary on both the short precession timescale $t_{\rm pre}\propto r^{5/2}$ \cite{1994PhRvD..49.6274A} and the longer radiation-reaction timescale $t_{\rm RR}\propto r^4$ \cite{1964PhRv..136.1224P}. The individual spin angles $\theta_1$, $\theta_2$ and $\Delta\Phi$ all vary on $t_{\rm pre}$ (the same is true for other quantities typically used to parametrize spin precession, like $\chi_{\rm p}$ from \cite{2014PhRvL.113o1101H,2015PhRvD..91b4043S}). Binaries undergo many precession cycles from formation to detection. This essentially randomizes the spin angle distribution (with some constraints, e.g. that $\chi_{\rm eff}$ must be constant \cite{2008PhRvD..78d4021R,2015PhRvD..92f4016G}). In other terms, the spin directions at detection are not indicative of the spin directions at formation. On the other hand, the spin morphology is an averaged quantity. It describes the \emph{shape} of the precession cones, not a particular position of the spins along those cones. The spin morphology does vary, however, on the longer radiation-reaction timescale $t_{\rm RR}$ over which GW emission dissipates energy and angular momentum. In general, binaries are mostly circulating at large separations and transition towards the two librating morphologies as the inspiral proceeds \cite{2015PhRvL.114h1103K,2015PhRvD..92f4016G,2018JPhCS.957a2014G}. The most notable property of the spin morphology is therefore that
\emph{morphology encodes details of BH spin precession, but it does not vary on the precession timescale.}

 \begin{figure*}[t!]
\includegraphics[width=0.98\textwidth]{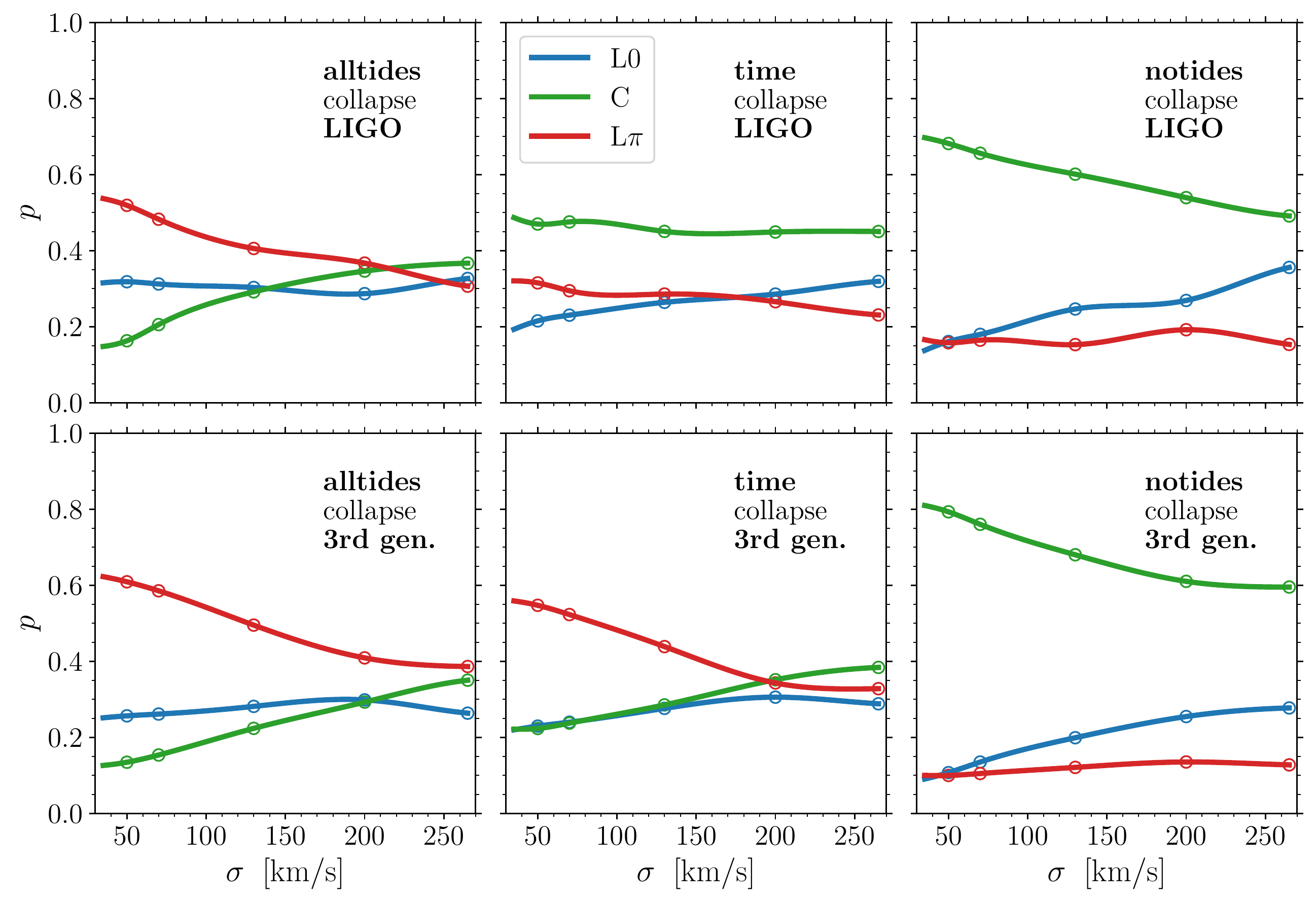}
\caption{Fraction $p$ of binaries in each of the three spin morphologies as a function of the kick speed $\sigma$. The three lines in each panel show the normalized detection rate for binaries which librate about $\Delta\Phi=0$ (L$0$, blue), circulate (C, green) or librate about $\Delta\Phi=\pi$ (L$\pi$, red) at the detection frequency of $f_{\rm GW}=20$ Hz. The three classes are mutually exclusive, i.e. $p({\rm L}0)+p({\rm C})+p({\rm L}\pi)=1$. Left, middle and right panels are produced with our three different assumptions on tidal interactions. Top and bottom panels assume detection rate for LIGO and Cosmic Explorer, respectively. All panels are obtained assuming the \emph{collapse} model for the spin magnitudes.}
\label{morphkicks}
\end{figure*}

\subsection{Fraction of binaries in each morphology}
 
Figure~\ref{morphkicks} shows the relative fraction of binaries in each of the three spin morphologies as a function of natal kicks\footnote{Being defined in terms of the azimuthal projections of the spins, the morphology is formally ill defined in the limit where $\sin\theta_1=0 $ and $\sin\theta_2 =0$. Results in this section therefore exclude all $\sigma=0$ km/s models, where binaries do not get tilted.} for different tidal alignment models and GW detectors. More circulating (librating) binaries are predicted if tides are inefficient (efficient).  This is especially true if natal kicks are moderately  small  ($\sigma\lesssim 200$ km/s). For larger values, the tilt angle distributions are closer to being isotropic, which in turn results in comparable fractions of binaries in each of the three morphologies (note for instance the \emph{alltides} cases in Fig.~\ref{morphkicks}, where all morphologies approach a probability of $p\sim1/3$ at large $\sigma$). 

In the \emph{alltides}  case, the L$\pi$ morphology dominates over L$0$, mainly because it corresponds to the much more frequent case where the primary BH forms first (``BH1 BH2'').  In the \emph{notides} models, on the other hand, more binaries are typically found in L$0$ than in L$\pi$ (but both are subdominant with respect to C). This is because the region of the parameter space which is influenced by the L$0$ transitions extends closer to the  $\theta_1\simeq \theta_2$ region, where \emph{notides} binaries form (cf. Fig.~14 in \cite{2015PhRvD..92f4016G}).

In the \emph{time} case, both librating and circulating binaries are present and track stellar progenitors that did (or did not) undergo tidal alignment. For the LIGO detector at design sensitivity, the fraction of detectable binaries in each morphology is roughly independent of $\sigma$. We find $p({\rm C})\simeq 0.5$ and $p({\rm L}0)\simeq p({\rm L}\pi)\simeq  0.25$. For third-generation interferometers, the detectable sample becomes largely dominated by L$\pi$ binaries. As we discuss below, the reason is that the likelihood to transition towards the two librating morphologies strongly depends on the binary's total mass.  
 
 \subsection{Spin morphology as a function of mass}

As already mentioned in Sec.~\ref{margdistchieff}, future detectors are expected to observe many systems with total mass $M\lesssim 30 M_\odot$ which are invisible to LIGO.
These systems have, on average, lower tidal timescales $t_{\tau}$ [cf. Eq.~(\ref{ttau})] and lower orbital velocities at the time of the first explosion, so they are more easily tilted by natal kicks and subsequently realigned by tides. These two effects preferentially populate regions of the $(\theta_1,\theta_2)$ plane far from the $\theta_1=\theta_2$ diagonal, which are more strongly affected by morphological transitions (see Fig.~\ref{deltaphisplit}). 
As illustrated in Fig. \ref{morphkicks}, instrumental improvements at high frequencies might dramatically change the expected number of sources in each morphology. For the \emph{time} model, we predict that a third-generation detector will observe more L$\pi$ than C binaries. The vast majority of these systems are low-mass binaries in the ``BH1 BH2'' channel subject to tidal realignment. 

\begin{figure}
\includegraphics[width=\columnwidth]{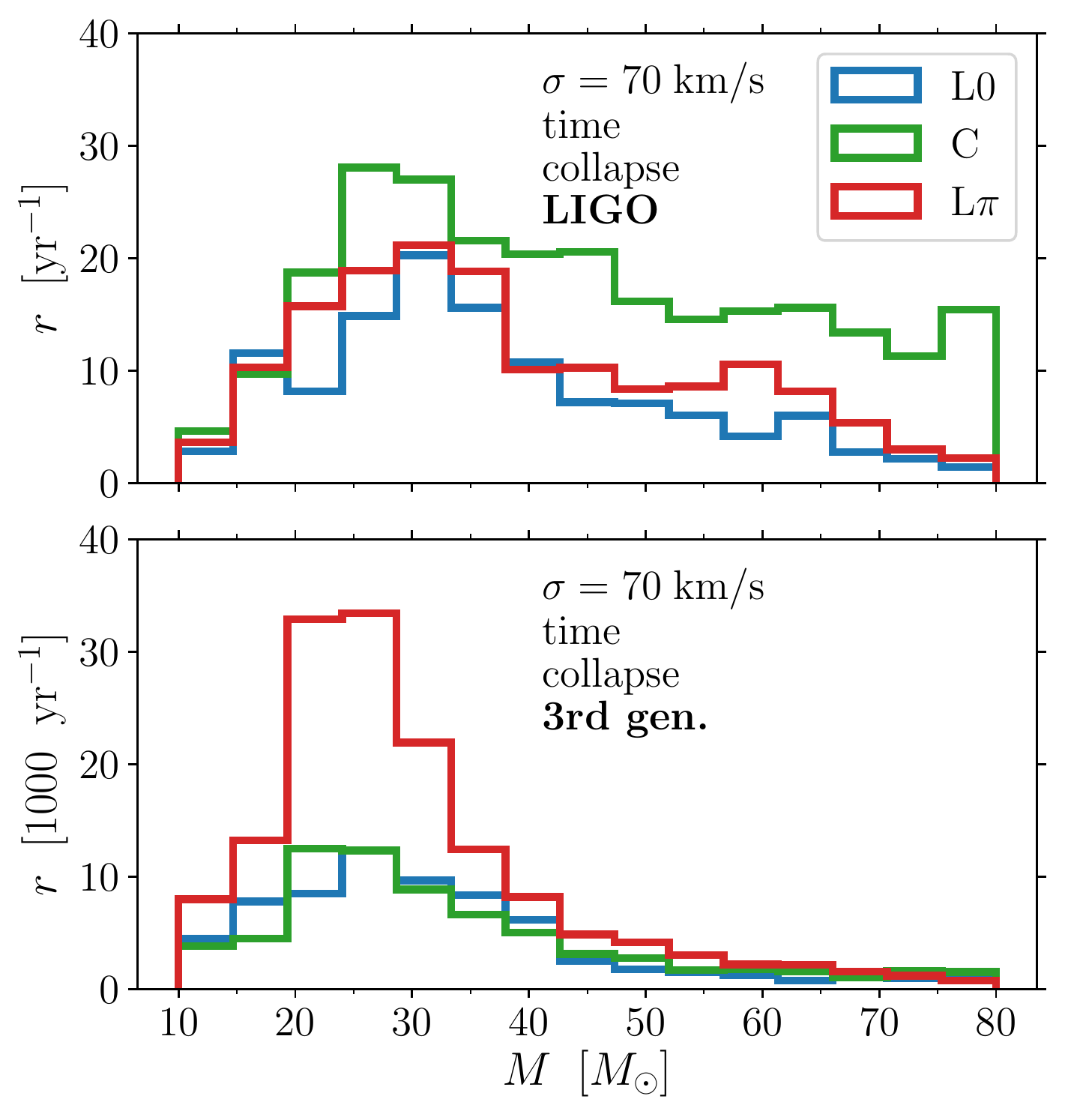}
\caption{LIGO detection rates $r$ for binary BHs in bins of total mass $M$, divided into the three spin morphologies. Low-mass systems subject to tidal realignment are mostly found in the L$\pi$ morphology. They are  invisible to LIGO but will dominate the detection rate for future detectors like Cosmic Explorer. Most binaries of high mass subject to pulsational pair-instability supernova are found in the C morphology.}
\label{morphmass}
\end{figure}
The mass dependence of the detectable fraction of binaries in each morphology is shown in Fig.~\ref{morphmass} for one of our \emph{time} models at moderate natal kicks ($\sigma=70$ km/s). We observe many low-mass, L$\pi$ binaries, which are invisible to current detectors but might dominate the rates in the third-generation era.
This is a strong effect, but it remains subdominant with respect to the assumptions on tidal interactions: a robust conclusion of our study is that binaries are preferentially librating (circulating) if tides are (in)efficient. 

Another  interesting feature can be seen  in Fig.~\ref{morphmass} at the high-mass end of the distributions ($M\sim 80 M_\odot$). For the same reason why low-mass binaries preferentially librate, high-mass binaries are mostly expected to circulate. Binaries with $M\sim 80 M_\odot$, which may be subject to pulsational pair-instability supernovae \cite{2016A&A...594A..97B}, are almost exclusively found in the circulating morphology. Our findings highlight a possible correlation between supernova physics and spin-precession dynamics, which deserves further investigation.

 \subsection{Dependence on the formation pathway}

\begin{figure*}[p!]
\includegraphics[page=1,width=0.47\textwidth]{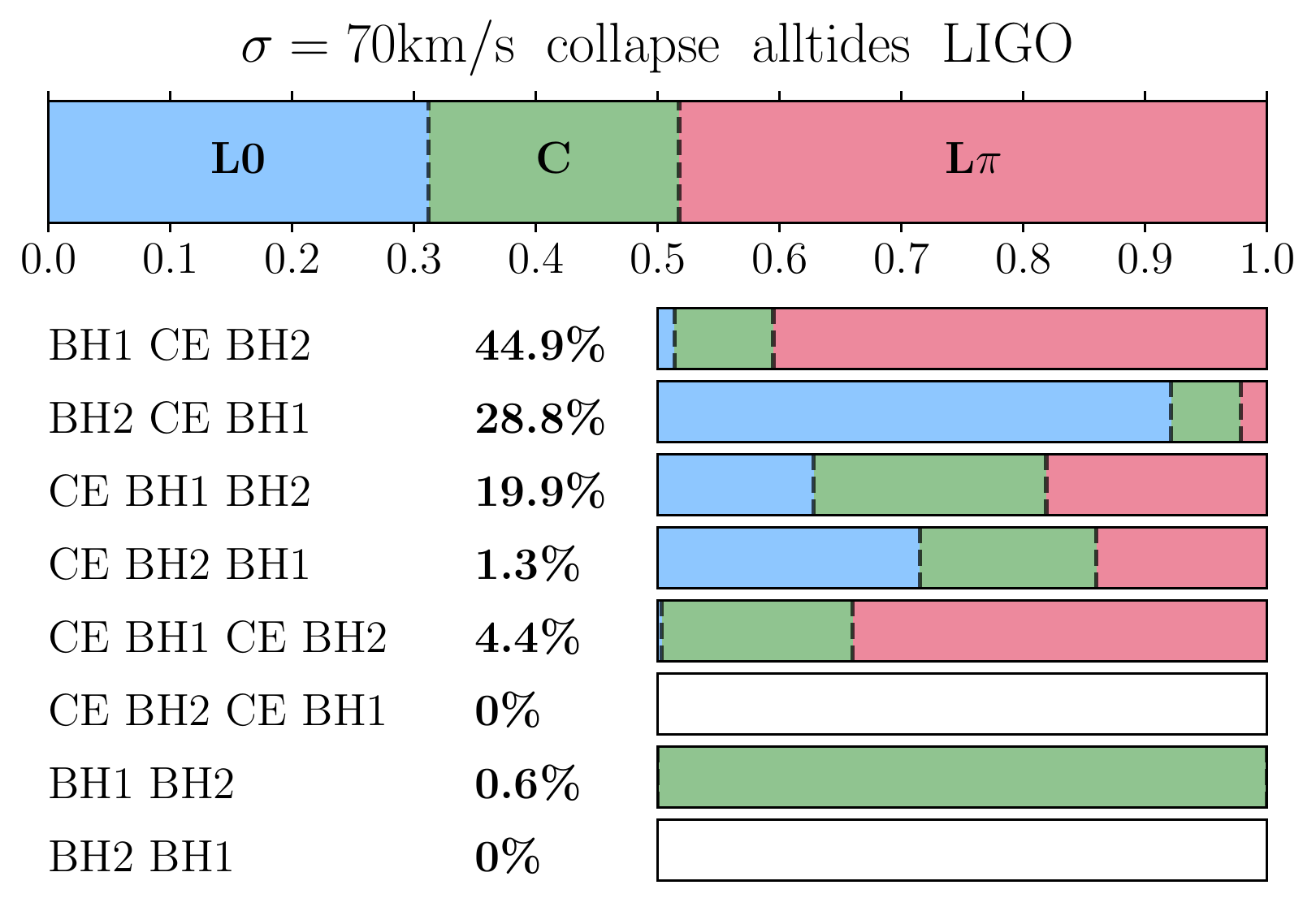}
\hspace{0.5cm}
\includegraphics[page=2,width=0.47\textwidth]{morphology_bars}\\
\vspace{0.5cm}
\includegraphics[page=3,width=0.47\textwidth]{morphology_bars}
\hspace{0.5cm}
\includegraphics[page=4,width=0.47\textwidth]{morphology_bars}\\
\vspace{0.5cm}
\includegraphics[page=5,width=0.47\textwidth]{morphology_bars}
\hspace{0.5cm}
\includegraphics[page=6,width=0.47\textwidth]{morphology_bars}

\caption{Bars showing the fractions of  binaries detectable by LIGO in each of the three spin-precession morphologies: librating about $\Delta\Phi=0$ (L$0$, blue), circulating (C, green) and librating about $\Delta\Phi=\pi$ (L$\pi$, red). Each panel is produced from a different population-synthesis simulation varying over assumptions on natal kicks (left: $\sigma=70$ km/s; right: $\sigma=200$ km/s) and  tidal interactions (top: \emph{alltides}; middle: \emph{time}; bottom: \emph{notides}); the spin-magnitude model is kept fixed to \emph{collapse}. The larger bar at the top of each panel shows the overall fraction of binaries in that particular simulation. The smaller bars instead, only consider binaries in each of the eight formation pathways of Sec.~\ref{pathway}, where ``BH1'' (``BH2'') stands for the formation of the more (less) massive BH and ``CE'' stands of a common-envelope phase. The percentage in boldface next to each small bar indicates the fraction of the LIGO detection rate originating from that particular subchannel.}
\label{morphologybars}
\end{figure*}

We now present a classification of the BH binaries in our samples based on both, the three spin morphologies and the eight formation pathways of Sec.~\ref{pathway}. This information is summarized in Fig.~\ref{morphologybars} for some indicative runs among our simulations.
 
As already stressed above, strong tidal interactions preferentially populate the two librating morphologies. This is especially true for binaries formed in the more standard channels where a common-envelope phase takes place between the two explosions. In the \emph{alltides} case, almost all binaries formed in the ``BH1 CE BH2'' (``BH2 CE BH1'') channel are found in the L$\pi$ (L0) morphology. As we discussed in Sec.~\ref{pathway} and Fig.~\ref{ratechannels}, these two channels  dominate the detection rate for $\sigma\lesssim 100$ km/s. Therefore, a larger fraction of librating binaries is present for these values of the kicks  (assuming tides are efficient at realigning spins). For higher kicks, the ``CE BH1 BH2'' channel becomes more important. Binaries from this channel have on average smaller misalignments (because they have been hardened by a common-envelope phase before the first explosion) and are thus closer to equipartition among the three morphologies.

Transitions into any of the two librating morphologies are much less likely in the \emph{notides} case.  In these models, the ``BH1 CE BH2'' and ``BH2 CE BH1'' channels mostly generate circulating binaries. The contribution to the two librating morphologies mainly comes from the ``CE BH1 BH2'' and ``CE BH2 BH1'' channels, where rough equipartition is reached. Since these channels dominate in the large-$\sigma$ regime, the morphological classification in the \emph{alltides} and \emph{notides} cases becomes more similar when BH kicks are large.

\section{Conclusions}
\label{conclusions}

After the first LIGO detections, it is becoming more widely accepted by the scientific community that BH spin orientations
are possibly the cleanest indicators of BH-binary formation channels. In particular, binaries formed in dynamical
interactions are predicted to have randomly distributed spins, while conventional wisdom asserts that the spins of binaries
formed in isolation are \emph{more or less} aligned. In this paper, we carefully distinguished between
spin alignment \emph{at BH-binary formation} and \emph{as observed in GWs}, and we quantified the expected degree of (mis)alignment for the first time. We studied an extensive set of astrophysical models, combining for the first time state-of-the-art stellar population synthesis (\textsc{StarTrack}  \cite{2008ApJS..174..223B}) and advanced PN evolution schemes (\textsc{precession} \cite{2016PhRvD..93l4066G}). We quantified the impact of several model parameters --namely the strength of natal kicks, the spin magnitude at formation and the efficiency of tidal alignment-- on the population of spinning BH binaries detectable by current and future ground-based GW interferometers.

Within the context of these models, we showed that future measured distributions of  effective spins \emph{alone} could break
the degeneracy between spin orientation and spin magnitude encoded in the very definition of  $\chi_{\rm eff}$.
We also confirmed previous claims that binaries formed in isolation cannot produce a symmetric $\chi_{\rm eff}$ distribution \cite{2017Natur.548..426F,2018ApJ...854L...9F}, although individual binaries can have $\chi_{\rm eff}<0$ (in contrast with some previous claims \cite{2016ApJ...832L...2R}).

The directions of the individual spins have not been confidently measured so far,\footnote{GW151226 data contains hints of a primary-BH misalignment in the range $25^\circ \lesssim \theta_1\lesssim 80^\circ$ \cite{2016PhRvL.116x1103A,2017PhRvL.119a1101O}.} but louder events, improved waveform models and more sophisticated parameter-estimation techniques may soon allow us to characterize the full (two-spin) dynamics of BH binaries. As shown here, this can have a significant payoff: we may be able to reconstruct the binary's formation history. Our study confirms some of our earlier results \cite{2013PhRvD..87j4028G}, and in particular the observation that the azimuthal precession phase $\Delta\Phi$ encodes clean information on processes that may (or may not) realign stellar spins in between the two core-collapse events forming each BH. We also presented the first prediction of how detectable sources would be distributed in terms of the recently discovered spin morphology \cite{2015PhRvL.114h1103K,2015PhRvD..92f4016G}, a feature of spin precession that does not vary on the precessional timescale.

In this paper, rather than focusing on fine model-parameter searches to reproduce current
LIGO/Virgo  observations, we have preferred to present only predictions from a limited set of astrophysically reasonable simulations.
Initial comparisons of our predictions with GW data \cite{2018PhRvD..97d3014W} found that observations from the first LIGO/Virgo observing run constrain $\sigma$ to be $\simeq 200$ ($\simeq 50$) km/s for (in)efficient tides, and marginally prefer small spin magnitudes.  Combining the formalism of \cite{2018PhRvD..97d3014W} and the more sophisticated predictions of this paper is an interesting avenue for future work. Furthermore, we plan to explore more advanced model selection techniques (e.g. \cite{2018arXiv180506442W,2018arXiv180608365T}) and to make detailed predictions for the next observing runs of the growing LIGO/Virgo/KAGRA network. 

Little information will be learned on processes affecting binary BH spins if their magnitudes turn out to be consistently very low. \emph{If} there is something out there to learn, however, the modeling efforts presented in this paper highlight the immense potential of future spin measurements. We are approaching the time when large GW detection catalogs will become available, and GW astronomy will turn into a large-statistics, data-driven field. With the rapid sensitivity improvements of ground-based interferometers, this may well happen sooner rather than later.
\pagebreak

\acknowledgments

We thank Christopher Berry, Sofia Maria Consonni, Jakub Klencki, 
Nathan Steinle and Colm Talbot for useful discussions and technical help. Data to reproduce results of this paper are publicly available at \href{https://github.com/dgerosa/spops}{github.com/dgerosa/spops} \cite{spops}.
D.G. is supported by NASA through Einstein Postdoctoral Fellowship Grant No. PF6-170152 awarded by the Chandra X-ray
Center,  operated by the Smithsonian Astrophysical Observatory for NASA under Contract NAS8-03060.
E.B. is supported by NSF Grants No. PHY-1841464 and No. AST-1841358, and by NSF-XSEDE Grant No. PHY-090003.
R.O.S. and D.W. gratefully acknowledge NSF Grant PHY-1707965.
K.B. acknowledges support from the Polish National Science Center (NCN) Grants Sonata Bis 2 DEC-2012/07/E/ST9/01360, No. LOFT/eXTP 2013/10/M/ST9/00729 and No. OPUS 2015/19/B/ST9/01099.
M. K. is supported by NSF Grant No. PHY-1607031. D.W. gratefully acknowledges support from the College of Science at Rochester Institute of Technology. This work has received funding from the European Union's Horizon 2020 research and innovation programme under the Marie Sk\l{}odowska-Curie grant agreement No. 690904.
Some of the computations were performed on the Caltech cluster \emph{Wheeler}, supported by the Sherman Fairchild Foundation and Caltech.

%\vspace{1cm}

\bibliography{stpn}
\end{document}